\documentclass[12pt]{article}
\usepackage{graphicx}
\usepackage{epstopdf}
\usepackage{subfig}
\usepackage{caption}
\usepackage{mathrsfs}
\usepackage{amssymb}
\usepackage{amsmath}
\usepackage{verbatim}
\advance\textheight by \topskip
\begin{document}

%\hoffset -0.15cm 
%\tolerance=100000
%\thispagestyle{empty}
%\setcounter{page}{1}

\newcommand{\HPA}[1]{{\it Helv.\ Phys.\ Acta.\ }{\bf #1}}
\newcommand{\AP}[1]{{\it Ann.\ Phys.\ }{\bf #1}}
\newcommand{\be}{\begin{equation}}
\newcommand{\ee}{\end{equation}}
\newcommand{\br}{\begin{eqnarray}}
\newcommand{\er}{\end{eqnarray}}
\newcommand{\ba}{\begin{array}}
\newcommand{\ea}{\end{array}}
\newcommand{\bi}{\begin{itemize}}
\newcommand{\ei}{\end{itemize}}
\newcommand{\bn}{\begin{enumerate}}
\newcommand{\en}{\end{enumerate}}
\newcommand{\bc}{\begin{center}}
\newcommand{\ec}{\end{center}}
\newcommand{\ul}{\underline}
\newcommand{\ol}{\overline}
\def\l{\left\langle}
\def\r{\right\rangle}
\def\as{\alpha_{s}}
\def\ycut{y_{\mbox{\tiny cut}}}
\def\yij{y_{ij}}
\def\epem{\ifmmode{e^+ e^-} \else{$e^+ e^-$} \fi}
\newcommand{\eeww}{$e^+e^-\rightarrow W^+ W^-$}
\newcommand{\qqQQ}{$q_1\bar q_2 Q_3\bar Q_4$}
\newcommand{\eeqqQQ}{$e^+e^-\rightarrow q_1\bar q_2 Q_3\bar Q_4$}
\newcommand{\eewwqqqq}{$e^+e^-\rightarrow W^+ W^-\ar q\bar q Q\bar Q$}
\newcommand{\eeqqgg}{$e^+e^-\rightarrow q\bar q gg$}
\newcommand{\eeqloop}{$e^+e^-\rightarrow q\bar q gg$ via loop of quarks}
\newcommand{\eeqqqq}{$e^+e^-\rightarrow q\bar q Q\bar Q$}
\newcommand{\eewwjjjj}{$e^+e^-\rightarrow W^+ W^-\rightarrow 4~{\rm{jet}}$}
\newcommand{\eeqqggjjjj}{$e^+e^-\rightarrow q\bar 
q gg\rightarrow 4~{\rm{jet}}$}
\newcommand{\eeqloopjjjj}{$e^+e^-\rightarrow q\bar 
q gg\rightarrow 4~{\rm{jet}}$ via loop of quarks}
\newcommand{\eeqqqqjjjj}{$e^+e^-\rightarrow q\bar q Q\bar Q\rightarrow
4~{\rm{jet}}$}
\newcommand{\eejjjj}{$e^+e^-\rightarrow 4~{\rm{jet}}$}
\newcommand{\jjjj}{$4~{\rm{jet}}$}
\newcommand{\qqbar}{$q\bar q$}
\newcommand{\ww}{$W^+W^-$}
\newcommand{\ar}{\rightarrow}
\newcommand{\sm}{${\cal {SM}}$}
\newcommand{\Dir}{\kern -6.4pt\Big{/}}
\newcommand{\Dirin}{\kern -10.4pt\Big{/}\kern 4.4pt}
\newcommand{\DDir}{\kern -8.0pt\Big{/}}
\newcommand{\DGir}{\kern -6.0pt\Big{/}}
\newcommand{\wwqqqq}{$W^+ W^-\ar q\bar q Q\bar Q$}
\newcommand{\qqgg}{$q\bar q gg$}
\newcommand{\qloop}{$q\bar q gg$ via loop of quarks}
\newcommand{\qqqq}{$q\bar q Q\bar Q$}

\def\st{\sigma_{\mbox{\scriptsize t}}}
\def\Ord{\buildrel{\scriptscriptstyle <}\over{\scriptscriptstyle\sim}}
\def\OOrd{\buildrel{\scriptscriptstyle >}\over{\scriptscriptstyle\sim}}
\def\jhep #1 #2 #3 {{JHEP} {\bf#1} (#2) #3}
\def\plb #1 #2 #3 {{Phys.~Lett.} {\bf B#1} (#2) #3}
\def\npb #1 #2 #3 {{Nucl.~Phys.} {\bf B#1} (#2) #3}
\def\epjc #1 #2 #3 {{Eur.~Phys.~J.} {\bf C#1} (#2) #3}
\def\zpc #1 #2 #3 {{Z.~Phys.} {\bf C#1} (#2) #3}
\def\jpg #1 #2 #3 {{J.~Phys.} {\bf G#1} (#2) #3}
\def\prd #1 #2 #3 {{Phys.~Rev.} {\bf D#1} (#2) #3}
\def\prep #1 #2 #3 {{Phys.~Rep.} {\bf#1} (#2) #3}
\def\prl #1 #2 #3 {{Phys.~Rev.~Lett.} {\bf#1} (#2) #3}
\def\mpl #1 #2 #3 {{Mod.~Phys.~Lett.} {\bf#1} (#2) #3}
\def\rmp #1 #2 #3 {{Rev. Mod. Phys.} {\bf#1} (#2) #3}
\def\cpc #1 #2 #3 {{Comp. Phys. Commun.} {\bf#1} (#2) #3}
\def\sjnp #1 #2 #3 {{Sov. J. Nucl. Phys.} {\bf#1} (#2) #3}
\def\xx #1 #2 #3 {{\bf#1}, (#2) #3}
\def\hepph #1 {{\tt hep-ph/#1}}
\def\preprint{{preprint}}

\def\beq{\begin{equation}}
\def\beeq{\begin{eqnarray}}
\def\eeq{\end{equation}}
\def\eeeq{\end{eqnarray}}
\def\a0{\bar\alpha_0}
\def\thrust{\mbox{T}}
\def\Thrust{\mathrm{\tiny T}}
\def\ae{\alpha_{\mbox{\scriptsize eff}}}
\def\ap{\bar\alpha_p}
\def\as{\alpha_{\mathrm{S}}}
\def\aem{\alpha_{\mathrm{EM}}}
\def\b0{\beta_0}
\def\cN{{\cal N}}
\def\cd{\chi^2/\mbox{d.o.f.}}
\def\Ecm{E_{\mbox{\scriptsize cm}}}
\def\ee{e^+e^-}
\def\enap{\mbox{e}}
\def\eps{\epsilon}
\def\ex{{\mbox{\scriptsize exp}}}
\def\GeV{\mbox{\rm{GeV}}}
\def\half{{\textstyle {1\over2}}}
\def\jet{{\mbox{\scriptsize jet}}}
\def\kij{k^2_{\bot ij}}
\def\kp{k_\perp}
\def\kps{k_\perp^2}
\def\kt{k_\bot}
\def\lms{\Lambda^{(n_{\rm f}=4)}_{\overline{\mathrm{MS}}}}
\def\mI{\mu_{\mathrm{I}}}
\def\mR{\mu_{\mathrm{R}}}
\def\MSbar{\overline{\mathrm{MS}}}
\def\mx{{\mbox{\scriptsize max}}}
\def\NP{{\mathrm{NP}}}
\def\pd{\partial}
\def\pt{{\mbox{\scriptsize pert}}}
\def\pw{{\mbox{\scriptsize pow}}}
\def\so{{\mbox{\scriptsize soft}}}
\def\st{\sigma_{\mbox{\scriptsize tot}}}
\def\ycut{y_{\mathrm{cut}}}
\def\slashchar#1{\setbox0=\hbox{$#1$}           % set a box for #1
     \dimen0=\wd0                                 % and get its size
     \setbox1=\hbox{/} \dimen1=\wd1               % get size of /
     \ifdim\dimen0>\dimen1                        % #1 is bigger
        \rlap{\hbox to \dimen0{\hfil/\hfil}}      % so center / in box
        #1                                        % and print #1
     \else                                        % / is bigger
        \rlap{\hbox to \dimen1{\hfil$#1$\hfil}}   % so center #1
        /                                         % and print /
     \fi}                                         %
\def\etmiss{\slashchar{E}^T}
\def\Meff{M_{\rm eff}}
\def\Ord{\lsim}
\def\OOrd{\gsim}
\def\tq{\tilde q}
\def\tchi{\tilde\chi}
\def\lsp{\tilde\chi_1^0}

\def\gam{\gamma}
\def\ph{\gamma}
\def\be{\begin{equation}}
\def\ee{\end{equation}}
\def\bea{\begin{eqnarray}}
\def\eea{\end{eqnarray}}
\def\lsim{\:\raisebox{-0.5ex}{$\stackrel{\textstyle<}{\sim}$}\:}
\def\gsim{\:\raisebox{-0.5ex}{$\stackrel{\textstyle>}{\sim}$}\:}

\def\ino{\mathaccent"7E} \def\gluino{\ino{g}} \def\mgluino{m_{\gluino}}
\def\sqk{\ino{q}} \def\sup{\ino{u}} \def\sdn{\ino{d}}
\def\chargino{\ino{\omega}} \def\neutralino{\ino{\chi}}
\def\cab{\ensuremath{C_{\alpha\beta}}} \def\proj{\ensuremath{\mathcal P}}
\def\dab{\delta_{\alpha\beta}}
\def\zz{s-M_Z^2+iM_Z\Gamma_Z} \def\zw{s-M_W^2+iM_W\Gamma_W}
\def\prop{\ensuremath{\mathcal G}} \def\ckm{\ensuremath{V_{\rm CKM}^2}}
\def\aem{\alpha_{\rm EM}} \def\stw{s_{2W}} \def\sttw{s_{2W}^2}
\def\nc{N_C} \def\cf{C_F} \def\ca{C_A}
\def\qcd{\textsc{Qcd}} \def\susy{supersymmetric} \def\mssm{\textsc{Mssm}}
\def\slash{/\kern -5pt} \def\stick{\rule[-0.2cm]{0cm}{0.6cm}}
\def\h{\hspace*{-0.3cm}}

\def\ims #1 {\ensuremath{M^2_{[#1]}}}
\def\tw{\tilde \chi^\pm}
\def\tz{\tilde \chi^0}
\def\tf{\tilde f}
\def\tl{\tilde l}
\def\ppb{p\bar{p}}
\def\gl{\tilde{g}}
\def\sq{\tilde{q}}
\def\sqb{{\tilde{q}}^*}
\def\qb{\bar{q}}
\def\sqL{\tilde{q}_{_L}}
\def\sqR{\tilde{q}_{_R}}
\def\ms{m_{\tilde q}}
\def\mg{m_{\tilde g}}
\def\Gs{\Gamma_{\tilde q}}
\def\Gg{\Gamma_{\tilde g}}
\def\md{m_{-}}
\def\eps{\varepsilon}
\def\Ce{C_\eps}
\def\dnq{\frac{d^nq}{(2\pi)^n}}
\def\DR{$\overline{DR}$\,\,}
\def\MS{$\overline{MS}$\,\,}
\def\DRm{\overline{DR}}
\def\MSm{\overline{MS}}
\def\ghat{\hat{g}_s}
\def\shat{\hat{s}}
\def\sihat{\hat{\sigma}}
\def\Li{\text{Li}_2}
\def\bs{\beta_{\sq}}
\def\xs{x_{\sq}}
\def\xsa{x_{1\sq}}
\def\xsb{x_{2\sq}}
\def\bg{\beta_{\gl}}
\def\xg{x_{\gl}}
\def\xga{x_{1\gl}}
\def\xgb{x_{2\gl}}
\def\lsp{\tilde{\chi}_1^0}

\def\gluino{\mathaccent"7E g}
\def\mgluino{m_{\gluino}}
\def\squark{\mathaccent"7E q}
\def\msquark{m_{\mathaccent"7E q}}
\def\M{ \overline{|\mathcal{M}|^2} }
\def\utm{ut-M_a^2M_b^2}
\def\MiLR{M_{i_{L,R}}}
\def\MiRL{M_{i_{R,L}}}
\def\MjLR{M_{j_{L,R}}}
\def\MjRL{M_{j_{R,L}}}
\def\tiLR{t_{i_{L,R}}}
\def\tiRL{t_{i_{R,L}}}
\def\tjLR{t_{j_{L,R}}}
\def\tjRL{t_{j_{R,L}}}
\def\tg{t_{\gluino}}
\def\uiLR{u_{i_{L,R}}}
\def\uiRL{u_{i_{R,L}}}
\def\ujLR{u_{j_{L,R}}}
\def\ujRL{u_{j_{R,L}}}
\def\ug{u_{\gluino}}
\def\utot{u \leftrightarrow t}
\def\ar{\to}
\def\sqk{\mathaccent"7E q}
\def\sup{\mathaccent"7E u}
\def\sdn{\mathaccent"7E d}
\def\chargino{\mathaccent"7E \chi}
\def\neutralino{\mathaccent"7E \chi}
\def\slepton{\mathaccent"7E l}
\def\M{ \overline{|\mathcal{M}|^2} }
\def\cab{\ensuremath{C_{\alpha\beta}}}
\def\ckm{\ensuremath{V_{\rm CKM}^2}}
\def\zz{s-M_Z^2+iM_Z\Gamma_Z}
\def\zw{s-M_W^2+iM_W\Gamma_W}
\def\s22w{s_{2W}^2}

\newcommand{\cpmtwo}    {\mbox{$ {\chi}^{\pm}_{2}                    $}}
\newcommand{\cpmone}    {\mbox{$ {\chi}^{\pm}_{1}                    $}}

\begin{flushright}
{SHEP-09-31}\\
\today
\end{flushright}
\vskip0.1cm\noindent
\begin{center}
{{\Large {\bf Tree Level Unitarity Bounds \\[0.25cm]
      for the Minimal $B-L$ Model}}
\\[1.0cm]
{\large L. Basso, A. Belyaev, S. Moretti and G. M. Pruna}\\[0.30 cm]
{\it  School of Physics and Astronomy, University of Southampton,}\\
{\it  Highfield, Southampton SO17 1BJ, UK.}
}
\\[1.25cm]
\end{center}

\begin{abstract}
{\small
\noindent
We have derived the unitarity bounds in the high energy limit for the
minimal $B-L$ extension of the Standard Model by analysing the full
class of Higgs and would-be Goldstone boson two-to-two scatterings at
tree level. Moreover, we have investigated how these limits could vary
at some lower critical value of the energy.
}

\end{abstract}

%\newpage

%%%%%%%%%%%%%%%%%%%%%%%%%%%%%%%%%%%%%%%%%%%%%%%%%%%%%%%%%%

\section{Introduction}
\label{Sec:Intro}
Despite there is no experimental evidence of a Higgs boson, the Higgs
mechanism is still largely considered as one of the preferred means of
generating masses for all known
(and possibly new) particles. For this reason, in the last three
decades or so, a global effort has been done to profile the massive
scalar boson(s) coming from Electro-Weak Symmetry Breaking (EWSB),
both within the Standard Model (SM) and Beyond it (BSM).

In the SM there is just one Higgs doublet consisting of four real
scalar fields, three of which, after spontaneous EWSB, turn out
to be absorbed in the longitudinal polarisation component of each of
the three weak gauge bosons, $W^{\pm}$ and $Z$, whilst the fourth one
gives the physical 
Higgs state $H$. Even if there are models in which the mass of this
particle is predicted, this is not generally possible within the SM
framework (or any of its non-supersymmetrical extensions encompassing
the Higgs mechanism), hence several theoretical methods have been
developed to constrain its value
(see \cite{DicusMathur}, \cite{LeeQuiggThacker}, \cite{DashenNeuberger}).
For example, to stay with the SM,
the pioneeristic work of \cite{LeeQuiggThacker} showed 
that, when $m_H$ is greater than a critical value $\simeq 1$ TeV
(known as unitarity bound), the elastic spherical wave describing the
scattering of
the longitudinally polarised vector bosons at very high energy
($\sqrt{s} \rightarrow \infty$) violates unitarity at tree level and
the theory stops to be valid from a perturbative point of view.
Moreover, it has also been shown that, well before the infinite energy
limit, the unitarity bound is violated already at some lower energy
critical value $\sqrt{s_c}>m_H$. The strongest such a bound, that we
can call ``critical energy unitarity bound'', is $m_H\simeq 1.09$ TeV
(again, for the SM, for example see \cite{MaalampiSirkkaVilja}).
%%%%%%%%%%%%%%%%%%%%%%%%%%%%%%%%%%%%%%%%%%%%%%%%%%%%%%%%%%%%%%%%5
%%%%%%%%%%%%%%%%%%%%%%%%%%%%%%%%%%%%%%%%%%%%%%%%%%%%%%%%%%%%%%%%
%%%%%%% 2) Also the authors should comment on the Higgs mass limits
%%%%%%% from the unitarity bound in the other models with extended
%%%%%%% Higgs sectors (doublet, triplet), which are available in the
%%%%%%% literature.
%%%%%%%%%%%%%%%%%%%%%%%%%%%%%%%%%%%%%%%%%%%%%%%%%%%%%%%%%%%%%%5
%%%%%%%%%%%%%%%%%%%%%%%%%%%%%%%%%%%%%%%%%%%%%%%%%%%%%%%%%%%%%%5

In the past, several efforts have been devoted to applying these
methodologies to a variety of models in order to extract any possible
information on their allowed parameter space. In particular, it has
been already applied to scenarios with extended scalar sectors yet
with same gauge structure as the SM, like those with additional
singlets (for example, see \cite{Cynolter:2004cq}), doublets (for
example, see \cite{MaalampiSirkkaVilja} and \cite{HuffelPocsik} for
non-Supersymmetric scenarios
and \cite{CasalbuoniDominiciFerruglioGatto} for Supersymmetric ones),
triplets (for example, see \cite{Aoki:2007ah}). It has also
been shown that this approach is successful with respect to $U(1)$
gauge group extensions of the SM (for example, for the case of $E_6$
superstring-inspired minimal $U(1)$ extensions,
see \cite{Robinett:1986nw}).

%%%%%%%%%%%%%%%%%%%%%%%%%%%%%%%%%%%%%%%%%%%%%%%%%%%%%%%%%%%%%%%
%%%%%%%%%%%%%%%%%%%%%%%%%%%%%%%%%%%%%%%%%%%%%%%%%%%%%%%%%%%%%%5
In the present work, we want to apply these two methods to the minimal
$B-L$ gauged extension of the SM \cite{B-L:rev}. The latter, with
respect to the 
SM, consists of a further $U(1)_{B-L}$ gauge group, three right-handed
neutrinos and an
additional Higgs boson generated through the $U(1)_{B-L}$ symmetry
breaking, responsible for giving mass to an additional $Z'$ gauge
boson. It is important to note that in this model the ${B-L}$
breaking can take place at the TeV scale, i.e., far below that of any
Grand Unified Theory (GUT). (This $B-L$ scenario therefore has
interesting phenomenological implications at present and future
colliders \cite{B-L:LHC}.) Hence,
if one wants to study the aforementioned unitarity 
constraints in this framework, the presence of two Higgs fields and
four massive vector bosons should be taken into account.
%%%%%%%%%%%%%%%%%%%%%%%%%%%%%%%%%%%%%%%%%%%%%%%%%%%%%%%%%%%%%%%%5
%%%%%%%%%%%%%%%%%%%%%%%%%%%%%%%%%%%%%%%%%%%%%%%%%%%%%%%%%%%%%%%%
%%%%%%% 1) Unitarity has been tested in various models, as mentioned
%%%%%%% in the paper. However, I recommend that the authors check the
%%%%%%% list of references to earlier works: e.g. the paper by
%%%%%%% R. Robinett (Phys Rev D34 (1986) 182) is not referred to,
%%%%%%% although it also deals with unitarity bounds when an
%%%%%%% additional Higgs singlet is included. The authors should
%%%%%%% comment this work and explain what are the differences in
%%%%%%% their research, and whether they agree with the earlier work.
%%%%%%%%%%%%%%%%%%%%%%%%%%%%%%%%%%%%%%%%%%%%%%%%%%%%%%%%%%%%%%5
%%%%%%%%%%%%%%%%%%%%%%%%%%%%%%%%%%%%%%%%%%%%%%%%%%%%%%%%%%%%%%5
As we will show in the remainder of the paper, the main difference
with respect to analogous treatments of minimal $U(1)$ extensions of the
SM (as done in \cite{Robinett:1986nw}) is that we are not considering
here the possibility of fixing any of the free parameters of the model
by exploiting GUT arguments.
%%%%%%%%%%%%%%%%%%%%%%%%%%%%%%%%%%%%%%%%%%%%%%%%%%%%%%%%%%%%%%%%
%%%%%%%%%%%%%%%%%%%%%%%%%%%%%%%%%%%%%%%%%%%%%%%%%%%%%%%%%%%%%%%5

This work is organised as follows: in the next section we describe
the model in its relevant (to this analysis) parts, in the following
one we show the theoretical methods adopted to constrain the
Higgs masses, in section \ref{Sec:Results} we present
our numerical results, then we conclude in section \ref{Sec:Conclusions}.

%%%%%%%%%%%%%%%%%%%%%%%%%%%%%%%%%%%%%%%%%%%%%%%%%%%%%%%%%%

\section{The scalar sector of the minimal $B-L$ model}
\label{Sec:Model}
The model under study is the so-called ``pure'' or ``minimal''
$B-L$ model (see \cite{B-L:LHC} for conventions and references) 
since it has vanishing mixing between the two $U(1)_{Y}$ 
and $U(1)_{B-L}$ gauge groups.
In the rest of this paper we refer to this model simply as the ``$B-L$
model''.  In this model the classical gauge invariant Lagrangian,
obeying the $SU(3)_C\times SU(2)_L\times U(1)_Y\times U(1)_{B-L}$
gauge symmetry, can be decomposed as:
\begin{equation}\label{L}
\mathscr{L}=\mathscr{L}_{YM} + \mathscr{L}_s + \mathscr{L}_f
+ \mathscr{L}_Y \, ,
\end{equation}
where $\mathscr{L}_{YM}$, $\mathscr{L}_s$, $\mathscr{L}_f$ and
$\mathscr{L}_Y$ are the Yang-Mills, scalar, fermionic and Yukawa
sectors, respectively.
Since it has been proven that perturbative unitarity violation at high 
energy occurs only in 
vector and Higgs bosons elastic scatterings, our 
interest is focused on the vector boson and scalar sectors. In
particular, as intimated in the previous section, it is well known
that in such processes the amplitude of the spherical partial wave can 
exceed the unit value.

In this connection then, we want to stress again that this model has
an extended gauge sector, with an additional
electrically neutral weak gauge boson, $Z'$, with respect to the SM. 
To realise the Higgs mechanism (breaking the $SU(2)_L\times U(1)_Y$ as
well as 
$U(1)_{B-L}$ symmetries) we must in turn introduce at least a complex
Higgs field $\chi$, which is a singlet state.

Now, following the BRS invariance (see \cite{BecchiRouetStora}), we
know that the amplitude for
emission or absorption of a `scalarly' polarised gauge
boson becomes equal to the amplitude 
for emission or absorption of the related Goldstone boson,
and, in the high energy limit ($s \gg m^2_{W^{\pm},Z,Z'}$) the
amplitude involving the (physical) longitudinal polarisation of
gauge bosons approaches the one involving the (unphysical) scalar one
(Equivalence Theorem, see \cite{ChanowitzGaillard85}).

Since it is the spherical partial wave of the former
that gives rise to unitarity violation, the analysis of the
perturbative unitarity of two-to-two
particle scattering in the gauge sector can be performed, in the high
energy limit, by exploiting the Goldstone sector.

Moreover, while evaluating scalar bosons scattering amplitudes, we
have explicitely verified by numerical computation that, in the search
for the Higgs mass limits, the contribution that arises from the
intermediate vector boson exchange is not relevant. Hence, in the high
energy limit, we can substitute the vector boson and Higgs boson
sectors with the related (would-be) Goldstone and Higgs boson sectors.

For the purpose of this work, we will therefore focus on the scalar
interacting Lagrangian of the Higgs and would-be Goldstone sectors (in the
Feynman gauge), i.e., the scalar Lagrangian neglecting the gauge couplings
in the covariant derivative.

The scalar Lagrangian is:
\begin{equation}\label{new-scalar_L}
\mathscr{L}_s=\left( D^{\mu} H\right) ^{\dagger} D_{\mu}H + 
\left( D^{\mu} \chi\right) ^{\dagger} D_{\mu}\chi - V(H,\chi ) \, ,
\end{equation}
{with the scalar potential given by}
\begin{equation}\label{new-potential}
V(H,\chi ) = -m^2H^{\dagger}H -
 \mu ^2\mid\chi\mid ^2 +
  \lambda _1 (H^{\dagger}H)^2 +\lambda _2 \mid\chi\mid ^4 + \lambda _3
 H^{\dagger}H\mid\chi\mid ^2  \, ,
\end{equation}
where $H$ and $\chi$ are the complex scalar Higgs doublet and singlet
fields, respectively:
\begin{eqnarray}
H=\frac{1}{\sqrt{2}}
\left(
\begin{array}{c}
-i(w^1-iw^2) \\
v+(h+iz)
\end{array}
\right), \qquad
\chi =
\frac{1}{\sqrt{2}}
(x+(h'+iz')),
\nonumber
\end{eqnarray}
where $w^{\pm}=w^1\mp iw^2$, $z$ and $z'$ are would-be Goldstone
bosons of $W^{\pm}$, $Z$ and $Z'$, respectively. Even for a minimal
$B-L$ model we have a generic mixing between $h$ and $h'$. Considering
$h_1$
and $h_2$ (with $m_{h_1}<m_{h_2}$) as the
two Higgs mass eigenstates corresponding to the two mass eigenvalues
\begin{eqnarray}
m^2_{h_1}&=& \lambda_1 v^2 + \lambda_2 x^2-\sqrt{(\lambda_2 x^2
- \lambda_1 v^2)^2+ (\lambda_3 x v)^2}, \\
m^2_{h_2}&=& \lambda_1 v^2 + \lambda_2 x^2 +\sqrt{(\lambda_2 x^2
- \lambda_1 v^2)^2+ (\lambda_3 x v)^2},
\end{eqnarray}
we can write the matrix that realises the Higgs mixing as
\begin{eqnarray}
\left(
\begin{array}{c}
h_1 \\
h_2
\end{array}
\right)=
\left(
\begin{array}{cc}
\cos{\alpha} & -\sin{\alpha} \\
\sin{\alpha} &\cos{\alpha}
\end{array}
\right)
\left(
\begin{array}{c}
h \\
h'
\end{array}
\right),
\end{eqnarray}
with $\alpha \in [-\frac{\pi}{2},\frac{\pi}{2}]$ (since the system is
invariant under $\alpha \rightarrow \alpha + \pi$ we halve the
domain of the orthogonal transformation), and
\begin{eqnarray}
\sin{(2\alpha)} &=& \frac{\lambda_3 x v}{\sqrt{(\lambda_2 x^2
- \lambda_1 v^2)^2 + (\lambda_3 x v)^2}}, \\
\cos{(2\alpha)} &=& \frac{ \lambda_2
x^2 - \lambda_1 v^2}{\sqrt{(\lambda_2 x^2-\lambda_1 v^2)^2 +
(\lambda_3 x v)^2}}.
\end{eqnarray}
In terms of physical parameters we can write
\begin{eqnarray}
v=\frac{m_W \sin{\theta_W}}{\sqrt{\pi \alpha}}=
\frac{m_W}{\sqrt{\pi \alpha_W}}, \qquad 
x=\frac{m_{Z'}}{2g'_1},
\end{eqnarray}
where $v(x)$ is the Higgs doublet(singlet) Vacuum Expectation Value
(VEV). 

From this, we can extrapolate the relation between $\lambda$'s and
Higgs masses and mixing angle:
\begin{eqnarray}\label{transformation}
\lambda_1&=&\frac{m_{h_1}^2}{2 v^2} + \frac{ \left(
m_{h_2}^2 - m_{h_1}^2 \right)}{2 v^2}\sin^2{\alpha} =
\frac{ m_{h_1}^2}{2v^2}\cos^2{\alpha} +  \frac{m_{h_2}^2}{2
v^2}\sin^2{\alpha}\nonumber \\
\lambda_2&=&\frac{m_{h_1}^2}{2 x^2} + \frac{ \left(
m_{h_2}^2 - m_{h_1}^2 \right)}{2 x^2}\cos^2{\alpha} =
\frac{ m_{h_1}^2}{2x^2}\sin^2{\alpha} +  \frac{m_{h_2}^2}{2
x^2}\cos^2{\alpha} \nonumber \\
\lambda_3&=&\frac{ \left(
m_{h_2}^2 - m_{h_1}^2 \right)}{ 2vx}\sin{(2\alpha)}
\end{eqnarray}
hence, we can calculate the explicit form of the interactions
(i.e., the Feynman rules) of the Lagrangian in terms of mass
eigenstates and couplings. We have listed the complete set of these
functions in appendix \ref{app:a}.

%%%%%%%%%%%%%%%%%%%%%%%%%%%%%%%%%%%%%%%%%%%%%%%%%%%%%%%%%%

\section{Theoretical bounds on the Higgs boson masses in the $B-L$ model}
\label{Sec:bounds}
In this section we want to explain in some detail the techniques that
we have used in order to obtain the aforementioned unitarity
bounds. The salient idea stems from the connection between
perturbative unitarity of a theory and a consequent upper bound on the
Higgs mass, or masses (in BSM scenarios), firstly described in detail
by \cite{LeeQuiggThacker}.

The well known result is that by evaluating the tree-level scattering
amplitude of longitudinally polarised vector bosons one finds that the
latter grows with the energy of the process, eventually violating
unitarity, unless one includes some other (model dependent)
interactions.

As already intimated, we also know that the equivalence theorem allows
one to
compute the amplitude of any process with external longitudinal vector
bosons $V_L$ ($V = W^\pm,Z,Z' $), in the limit $m^2_V\ll s$,
by substituting each one of them with the related Goldstone bosons $v
= w^\pm,z,z'$ and its general validity is proven
(see \cite{ChanowitzGaillard85}); schematically, if we consider a
process with four longitudinal vector bosons: $M(V_L V_L \rightarrow
V_L V_L) = M(v v \rightarrow v v)+ O(m_V^2/s)$.

We have also verified %in Appendix \ref{app:b}
that, in the high energy limit, the
intermediate vector boson exchange does not play a fundamental role in
the Higgs boson(s) limits search, hence we simplify our approach by
employing a theory of interacting would-be Goldstone
bosons $v = w^\pm,z,z'$ described by the scalar Lagrangian in
eq.~(\ref{new-scalar_L}).

We therefore studied the unitarity constraints in the $B-L$ model by
calculating tree-level amplitudes for all two-to-two processes 
involving the full set of possible (pseudo)scalar fields (the most
relevant subset is given by table \ref{tab:channels}).

Given a tree-level scattering amplitude between two spin-$0$ particles,
$M(s,\theta)$, where $\theta$ is the scattering (polar) angle, 
we know that the partial wave amplitude with angular
momentum $J$ is given by
\begin{eqnarray}\label{integral}
a_J = \frac{1}{32\pi} \int_{-1}^{1} d(\cos{\theta}) P_J(\cos{\theta})
M(s,\theta),
\end{eqnarray}
where $P_J$ are Legendre polynomials. It has been proven
(see \cite{LuscherWeisz}) that, in order to preserve unitarity, each
partial wave must be bounded by the condition
\begin{eqnarray}\label{condition}
|\textrm{Re}(a_J(s))|\leq \frac{1}{2}.
\end{eqnarray}

It turns out that only $J=0$ (corresponding to the spherical partial
wave contribution) leads to some bound, so we will not discuss the
higher partial waves any further. 

We have verified that, in the high energy limit, only the four-point
vertexes (related to the four-point functions of the interacting
potential, eqs.~(\ref{4-goldstone})--(\ref{4-higgs}) of
App.~\ref{app:a})
contribute to the $J=0$ partial wave amplitudes, and this
is consistent with many other works that exploit the same methodology
(for
example, 
see \cite{MaalampiSirkkaVilja},
\cite{HuffelPocsik} and \cite{CasalbuoniDominiciFerruglioGatto}). 

Hence, we will present the main results of our study focusing only on
the relevant subset of all
spherical partial wave amplitudes that is shown in
table \ref{tab:channels}. Here, we should notice that,
as one can conclude from direct computation, in the high energy limit
the contributions in table \ref{tab:channels} ticked
with $\sim$ are just a double counting of the channels ticked with
$\surd$ or combinations of them.

\begin{table}[!htbp]
\begin{center}
\begin{tabular}{|c|c|c|c|c|c|c|}
\hline
\  & $zz$ & $w^+w^-$ & $z'z'$ & $h_1h_1$ & $h_1h_2$ & $h_2h_2$ \\
\hline
$zz$ & $\surd$ & $\surd$ & $\surd$ & $\surd$ & $\surd$ &  $\surd$ \\
\hline
$w^+w^-$ & $\sim$ & $\surd$ & $\sim$ & $\sim$ & $\sim$ & $\sim$ \\
\hline
$z'z'$ & $\sim$ & $\sim$ & $\surd$ & $\surd$ & $\surd$ & $\surd$ \\
\hline
$h_1h_1$ & $\sim$ & $\sim$ & $\sim$ & $\surd$ & $\surd$ & $\surd$ \\
\hline
$h_1h_2$ & $\sim$ & $\sim$ & $\sim$ & $\sim$ & $\surd$ & $\surd$ \\
\hline
$h_2h_2$ & $\sim$ & $\sim$ & $\sim$ & $\sim$ & $\sim$ & $\surd$ \\
\hline
\end{tabular}
\end{center}
\caption{The most relevant subset of two-to-two scattering processes
in the minimal $B-L$ model in the Higgs and would-be Goldstone boson
sectors. The rows(columns) refer to
the initial(final) state (or vice versa).
The symbol $\sim$ refers to processes that can be computed by appropriate
rearrangements of those symbolised by $\surd$.}
\label{tab:channels}
\end{table}

Moreover, we have explicitely verified by numerical computation that
the main contributions come 
from the so-called scattering eigenchannels, i.e., the diagonal
elements of the ``matrix'' in table \ref{tab:channels}. In particular,
for our choice of method,
only the $zz \to zz$ and $z'z' \to z'z'$,
and to a somewhat lesser extent also $h_1h_1\to h_1h_1$ and 
$h_2h_2\to h_2h_2$,
play a relevant role. For
completeness, we list here all the $a_0$'s, 
eigenchannel by eigenchannel\footnote{Actually, in the high energy
limit, the $a_0(w^+w^-\rightarrow w^+w^-)$ differs from
eq.~(\ref{a0ww}) by a quantity $\simeq \alpha_W$
where photon and $Z$-boson exchange in
the $t$-channel, but since we are applying the condition in
eq.~(\ref{condition}) and $\alpha_W \ll \frac{1}{2}$, this correction
does not change the picture of our Higgs boson mass limit search.}:
\begin{eqnarray}\label{a0zz}
a_0(zz\rightarrow zz)
&=& \frac{3 \alpha_W}{32
m_W^2}  \left[ m_{h_1}^2 + m_{h_2}^2
+ \left( m_{h_1}^2 -
m_{h_2}^2 \right) \cos{(2\alpha)} \right],
\end{eqnarray}
\begin{eqnarray}\label{a0ww}
 a_0(w^+w^-\rightarrow w^+w^-) &=&
\frac{\alpha_W}{16 m_W^2} \left[ m_{h_1}^2 + m_{h_2}^2 + \left( m_{h_1}^2
- m_{h_2}^2 \right) \cos{(2\alpha)} \right],
\end{eqnarray}
\begin{eqnarray}\label{a0zpzp}
a_0(z'z'\rightarrow z'z') &=&
\frac{3}{32 \pi  x^2} \left[ m_{h_1}^2 + m_{h_2}^2 - \left( m_{h_1}^2
- m_{h_2}^2 \right) \cos{(2\alpha)} \right],
\end{eqnarray}
\begin{eqnarray}\label{a0h1h1}
a_0(h_1h_1\rightarrow h_1h_1) &=&
\frac{3\alpha_W}{32 m_W^2}
\left[ m_{h_1}^2 + m_{h_2}^2 + \left( m_{h_1}^2 -
m_{h_2}^2 \right) \cos{(2\alpha)} \right] \cos^4{\alpha}  \nonumber \\
&-&
\frac{3\sqrt{\alpha_W}}{64 m_W \sqrt{\pi} x}
\left( m_{h_1}^2 -
m_{h_2}^2 \right) \sin^3{(2\alpha)}  \nonumber \\
&+&
\frac{3}{16  \pi  x^2}\left[
m_{h_1}^2  -
\left( m_{h_1}^2 -
m_{h_2}^2 \right) \cos^2{\alpha} \right] \sin^4{\alpha},
\end{eqnarray}
\begin{eqnarray}\label{a0h1h2}
a_0(h_1h_2\rightarrow h_1h_2) &=&
\frac{ \sqrt{\alpha_W }}{256 m_W \sqrt{\pi} x}
\left( m_{h_1}^2 - m_{h_2}^2 \right) ( \sin{(2\alpha)} -
3 \sin{(6\alpha)})  \nonumber \\
&+&
\frac{3}{64 \pi x^2}
\left[ m_{h_1}^2 - \left( m_{h_1}^2 -
m_{h_2}^2 \right) \cos^2{\alpha} \right] \sin^2{(2\alpha)} \nonumber \\
&+&
\frac{3 \alpha_W}{64 m_W^2}
\left[ m_{h_1}^2 - \left( m_{h_1}^2 -
m_{h_2}^2 \right) \sin^2{\alpha} \right]
\sin^2{(2\alpha)},
\end{eqnarray}
\begin{eqnarray}\label{a0h2h2}
a_0(h_2h_2\rightarrow h_2h_2)&=&
\frac{3}{16 \pi x^2}
 \left[ m_{h_1}^2 - \left( m_{h_1}^2 -
 m_{h_2}^2 \right) \cos^2{\alpha} \right] \cos^4{\alpha} \nonumber \\
&-&
\frac{3  \sqrt{\alpha_W} }{64 m_W \sqrt{\pi } x} \left( m_{h_1}^2
- m_{h_2}^2 \right) \sin^3{(2\alpha)} \nonumber \\
&+&
\frac{3 \alpha_W}{16 m_W^2} \left[ m_{h_1}^2 - \left( m_{h_1}^2 -
m_{h_2}^2 \right) \sin^2{\alpha} \right] \sin^4{\alpha}.
\end{eqnarray}

We remark upon the fact that in the high energy limit, $\sqrt
s\rightarrow \infty$, only the $a_0$ partial wave amplitude (i.e., the
four-point function as one can conclude by direct comparison between
eqs.~(\ref{a0zz})--(\ref{a0h2h2})
and eqs.~(\ref{4-goldstone})--(\ref{4-higgs}) in App.~\ref{app:a}) does not
vanish, instead it approaches a value depending only on $m_{h_1}$,
$m_{h_2}$ and $\alpha$. Therefore, by applying the condition in
eq.~(\ref{condition}), we can obtain several different (correlated)
constraints on the Higgs masses and mixing angle, i.e., we can find the
$m_{h_1}$-$m_{h_2}$-$\alpha$ subspace in which the perturbative
unitarity of the theory is valid up to any energy scale.

Let us now come to another bound,
often referred to as ``triviality bound'' in some literature,
but which is essentially a unitarity bound obtained
at some finite energy. Another question one could ask is how
much  the above constraint would be relaxed if the analysis were done
not in the infinite energy limit $\sqrt s\rightarrow \infty$ but
rather at a critical energy value $\sqrt s\rightarrow \sqrt s_c$ for
which the bound on the Higgs masses is the most relaxed possible.

To this end, we developed a simple technique that is based on the
following idea: after fixing the Higgs mixing angle, we perform the
integral in eq.~(\ref{integral}) 
on the $m_{h_1}$-$m_{h_2}$ subspace defined by
eq.~(\ref{condition}) applied to the $a_0$'s by varying the value of
$\sqrt s$ in order to find the critical energy $\sqrt s_c$ for which
the result is maximal.

For illustration, let us consider the channels $zz\rightarrow zz$ and
$z'z'\rightarrow z'z'$ as we have already mentioned that they are the
most relevant ones. We
evaluated that their spherical partial waves, in the limit
$m_Z\ll m_{h_1}, m_{h_2}$ and $m_{Z'}\ll m_{h_1}, m_{h_2}$
respectively, are

\begin{eqnarray}\label{Ta0zz}
a_0(zz\rightarrow zz; s) &=& a_0(zz\rightarrow zz;
s\rightarrow \infty) \nonumber \\  &+&
\frac{\alpha_W m_{h_1}^2}{16 m_{W}^2} \left( \frac{m_{h_1}^2}{s -
m_{h_1}^2}
- \frac{2m_{h_1}^2}{s} \ln{\left[ \frac{\left( m_{h_1}^2 +
s \right)}{m_{h_1}^2} \right]} \right) \cos^2{\alpha} \nonumber \\
&+&
\frac{\alpha_W m_{h_2}^2}{16 m_{W}^2}
\left( \frac{m_{h_2}^2}{s -
m_{h_2}^2}
- \frac{2m_{h_2}^2}{s} \ln{\left[ \frac{\left( m_{h_2}^2 +
s \right)}{m_{h_2}^2} \right]} \right) \sin^2{\alpha},
\end{eqnarray}
\begin{eqnarray}\label{Ta0zpzp}
a_0(z'z'\rightarrow z'z'; s) &=& a_0(z'z'\rightarrow z'z';
s\rightarrow \infty) \nonumber \\  &+&
\frac{m_{h_1}^2}{16 \pi x^2} \left( \frac{m_{h_1}^2}{s -
m_{h_1}^2}
- \frac{2m_{h_1}^2}{s} \ln{\left[ \frac{\left( m_{h_1}^2 +
s \right)}{m_{h_1}^2} \right]} \right) \sin^2{\alpha} \nonumber \\
&+&
\frac{m_{h_2}^2}{16 \pi x^2}
\left( \frac{m_{h_2}^2}{s -
m_{h_2}^2}
- \frac{2m_{h_2}^2}{s} \ln{\left[ \frac{\left( m_{h_2}^2 +
s \right)}{m_{h_2}^2} \right]} \right) \cos^2{\alpha}.
\end{eqnarray}
We can integrate these functions then at fixed angle $\alpha$ in order
to establish which is the value of $\sqrt s$ that maximises the
integral (and thus the relative allowed configuration space), in turn
producing the most relaxed upper bound on the Higgs masses $m_{h_1}$
and $m_{h_2}$ at fixed $\alpha$.

In the next section we will explore the potential of these two
criteria, i.e., unitarity at either infinite or finite energy,
 in constraining the Higgs parameter space, in terms of the
mixing angle between the two physical Higgs fields and their masses.

%%%%%%%%%%%%%%%%%%%%%%%%%%%%%%%%%%%%%%%%%%%%%%%%%%%%%%%%%%

\section{Results}
\label{Sec:Results}
In the following subsections we will show that the most relevant
scattering channels for the unitarity analysis are pure-$z$ and pure
$z'$-bosons scatterings. As one can see from
eqs.~(\ref{a0zz})-(\ref{a0zpzp}), and
eqs.~(\ref{Ta0zz})-(\ref{Ta0zpzp}), the limit coming from these two 
channels is unaffected by the transformation $\alpha \rightarrow
-\alpha$, hence it is not restrictive to consider the half domain
$\alpha \in [0,\frac{\pi}{2}]$ only.
Furthermore, we remind the reader that we are still not allowing the
inversion of the Higgs mass eigenvalues, i.e., we still
require $m_{h_1}<m_{h_2}$.

\subsection{Unitarity bounds}

In this subsection we study the space of the parameters $\alpha$,
$m_{h_1}$ and $m_{h_2}$, once it has been specified by the unitarity
condition applied to the spherical partial wave scattering amplitudes
listed in the previous section in the very high energy limit.

For a start, let us mention that, after performing a  complete
numerical analysis, we discovered that there only two eigenchannels
that play any role in this study and these are the $zz\rightarrow zz$
and $z'z'\rightarrow z'z'$ scattering processes. Nevertheless, we must
also point out the fact that, for particular choices of $\alpha$, the
Higgs scattering channels (chiefly, $h_1h_1\to h_1h_1$ and $h_2h_2\to
h_2h_2$) partially bound the Higgs mass space just as the gauge boson
scattering channels. For example, if we choose $\alpha \rightarrow 0$,
then we have that the upper bound on the $m_{h_1}$ mass from
$h_1h_1\rightarrow h_1h_1$ is exactly the same as the one that we can
extract from the evaluation of $zz \rightarrow zz$, while if we choose
$\alpha \rightarrow \pi/2$, then we have that the upper bound on the
$m_{h_2}$ mass from $h_2h_2\rightarrow h_2h_2$ is exactly the same as
the one that we can extract from the evaluation of $z'z' \rightarrow
z'z'$.

We want to start our analysis in the $m_{h_1}$-$m_{h_2}$
subspace, hence we ``slice'' the $3$-dimensional parameter space we
are dealing with by
keeping the Higgs mixing angle fixed.
\begin{figure}[!ht]
  \subfloat[]{
  \label{a001}
  \includegraphics[angle=0,width=0.49\textwidth]{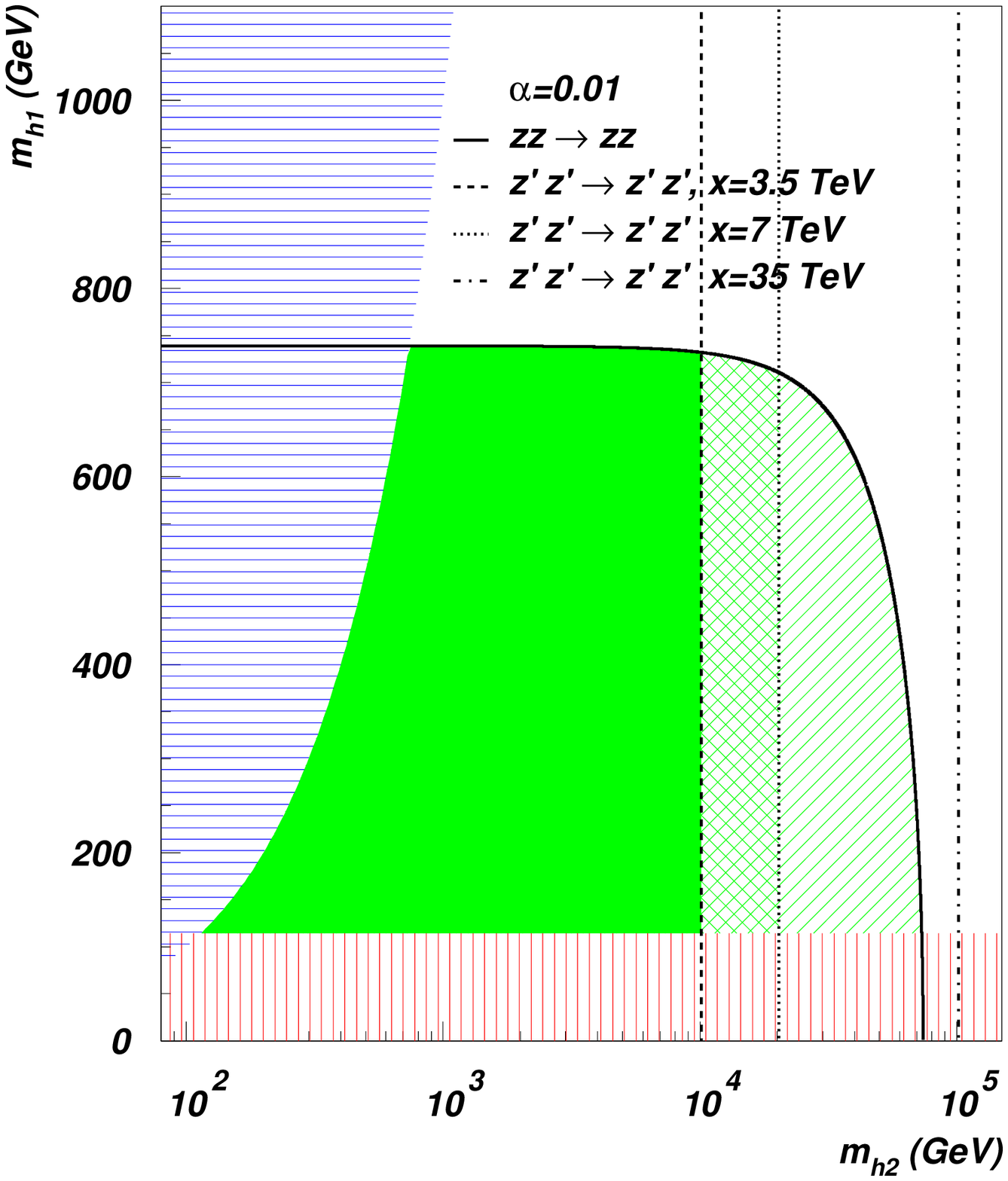}}
  \subfloat[]{
  \label{a01}
  \includegraphics[angle=0,width=0.49\textwidth ]{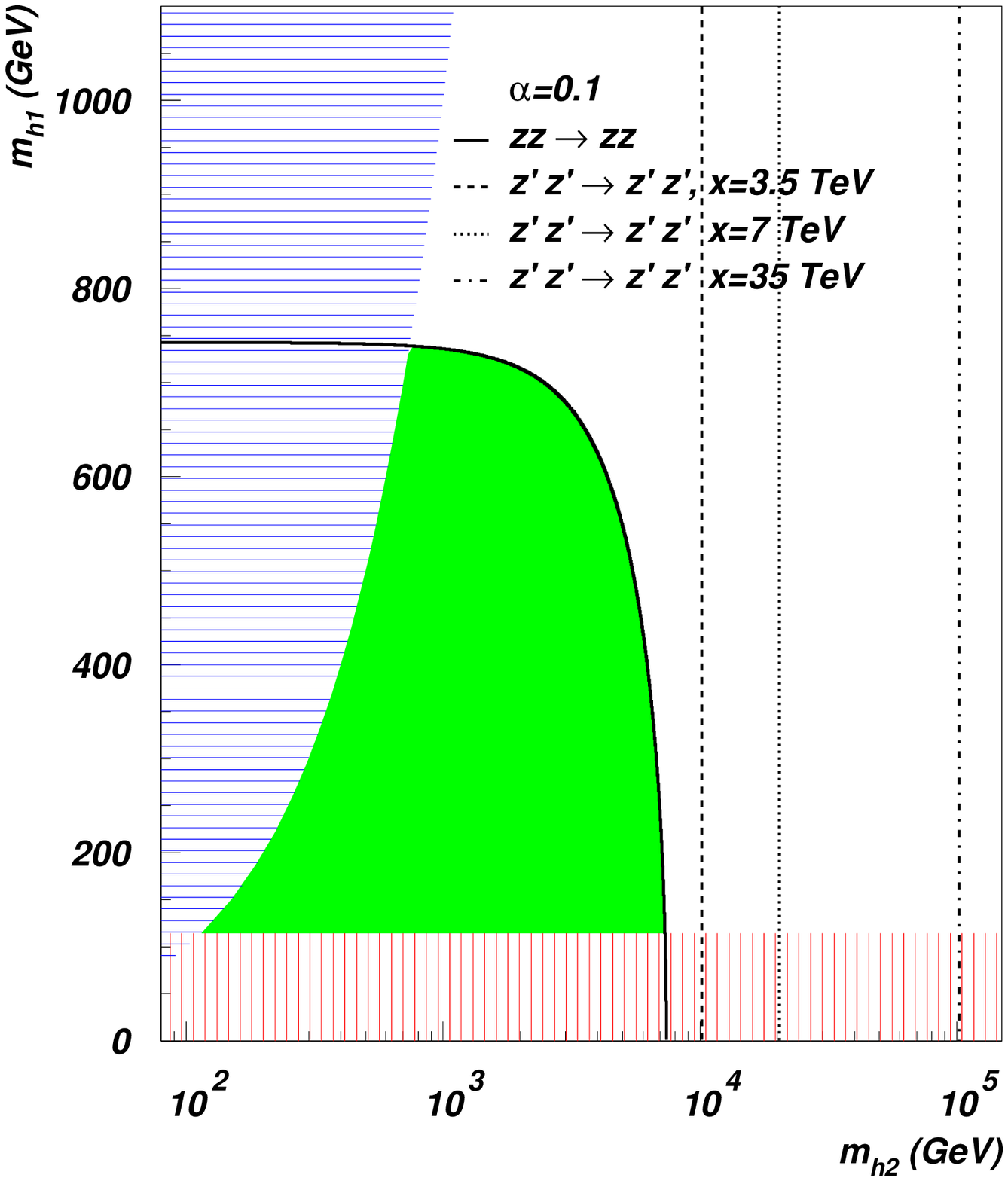}}
\\
  \subfloat[]{
  \label{api4}
  \includegraphics[angle=0,width=0.49\textwidth]{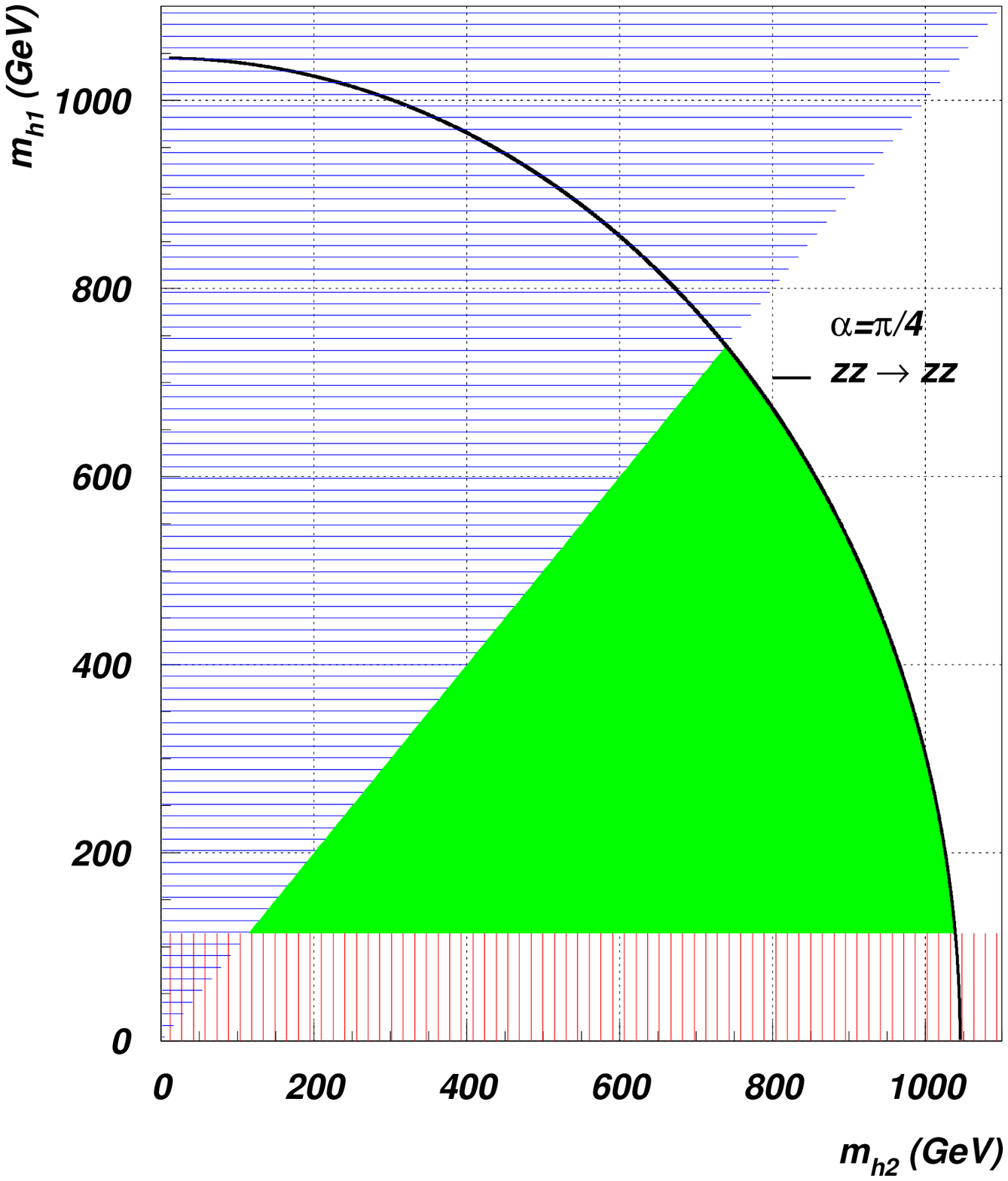}}
  \subfloat[]{
  \label{api2}
  \includegraphics[angle=0,width=0.49\textwidth]{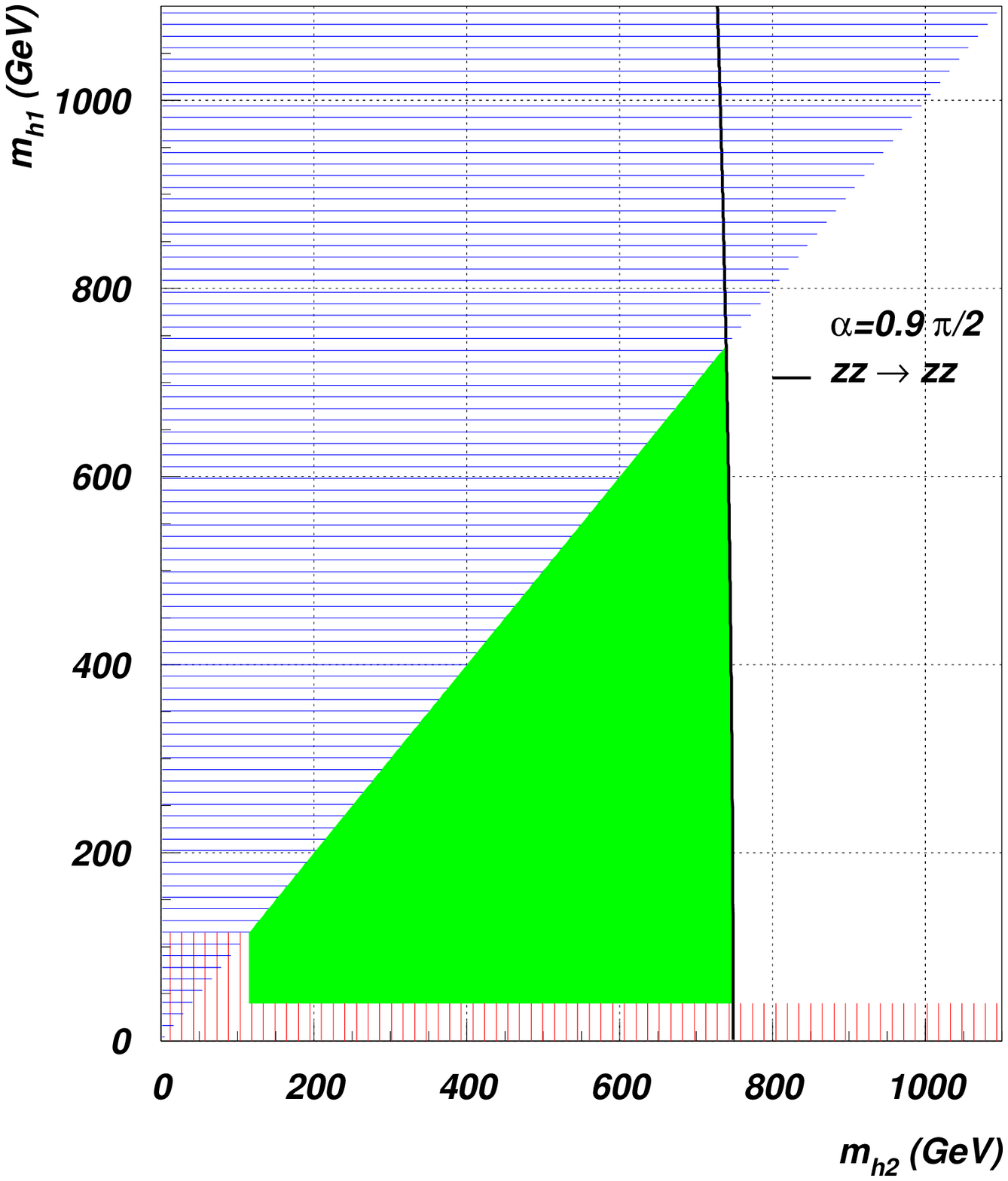}}
  \vspace*{-0.5cm}
  \caption{Higgs bosons mass limits in the $B-L$ model coming from
  the unitarity condition $|\textrm{Re}(a_0)|\le \frac{1}{2}$ applied
  to the $zz \rightarrow zz$ and $z'z' \rightarrow z'z'$ scatterings for
  several values of $x$, for $\alpha=0.01$ (\ref{a001}), $\alpha=0.1$
  (\ref{a01}), $\alpha=\pi/4$ (\ref{api4}) and $\alpha=0.9\ \pi/2$
  (\ref{api2}). The (blue) horizontal shadowed region corresponds to the
  unphysical configuration $m_{h_1}>m_{h_2}$. The (red) vertical
  shadowed region is excluded by the LEP experiments.}
  \label{mh1vsmh2}
\end{figure}
By applying the unitarity constraint to the spherical partial waves
listed in the previous section, one discovers that for a mixing angle
$\alpha$ such that 
\begin{eqnarray}
{\rm{arctan}}\left( \frac{ m_W}{x\sqrt{\pi \alpha_W }} \right)
\le \alpha \le \frac{\pi}{2}
\end{eqnarray}
the allowed parameter space is completely defined by the
$zz\rightarrow zz$ eigenchannel.

We will call ``high-mixing domain'' the parameter space defined by a
choice of the mixing angle in this range, while the ``low-mixing
domain'' is the complementary one. For example, since $x \geq 3.5$ TeV
following the LEP analyses \cite{Cacciapaglia:2006pk}, if we choose
$x$ to be exactly $3.5$ TeV, then we say that the high-mixing
domain, in this particular case, is the one for $ 0.073 \leq \alpha
\leq \frac{\pi}{2} $ (and, conversely, the low-mixing one is in the
interval $ 0 \leq \alpha < 0.073 $).

We can appreciate how the size of the Higgs mixing affects the limits
on the Higgs masses by looking at figure \ref{mh1vsmh2}, in which we
plot the allowed space for the latter, limitedly to the two
eigenchannels $zz \rightarrow zz$ and $z'z' \rightarrow z'z'$, for
four different values of $\alpha$ and three of $x$ (the latter affects
only the limit coming from the $z'z' \to z'z'$ scattering).

We see that in both cases, as expected, the light Higgs mass upper
bound does not exceed the SM one (which is $\simeq 700$ GeV, according
to~\cite{LuscherWeisz}),
%\footnote{Ours actually differs somewhat from
%  the value in \cite{LeeQuiggThacker} because we used the
%  $|\textrm{Re}(a_0)|\le \frac{1}{2}$ constraint instead of the one
%  they exploited, $|a_0|\le 1$, and the former has been shown to give
%  a more stringent condition than the one obtained from the
%  latter. Notice however that ref.~\cite{LeeQuiggThacker} worked this
%  limit out also exploiting isospin, instead of just using an
%  eigenchannel scattering argument as we are doing here.}
and it runs to the experimental lower limit from LEP (according
to \cite{Barate:2003sz}) as the
heavy Higgs mass increases. This is because the
two Higgses `cooperate' in the unitarisation of the eigenchannels so
that, if one Higgs mass tends to grow, the other one must become
lighter and lighter in order to keep the scattering matrix elements
unitarised.

While we are in the high-mixing domain, as in
figure \ref{a01}-\ref{api4}-\ref{api2} (where
$\alpha=0.1$, $\alpha=\frac{\pi}{4}$,
$\alpha=0.9\ \frac{\pi}{2}$, respectively\footnote{For the last of
these values of the mixing angle, the lower limit from LEP experiments
on the light Higgs boson is $m_{h_1}>40$ GeV, while for the firsts it
is almost equal to the SM lower limit ($m_{h_1}>115$ GeV)
as illustrated in figure \ref{mh1vsmh2}.}), the
allowed region coming from the $zz\to zz$
scattering is completely included in the $z'z'\to z'z'$ allowed area,
and the highest value allowed for the heavy Higgs mass only depends
on the mixing angle via
\begin{eqnarray}\label{high-maxmh2}
{\rm{Max}}(m_{h_2})=2\sqrt{\frac{2}{3}}
\ \frac{m_W}{\sqrt{\alpha_W}\sin{\alpha}}.
\end{eqnarray}

When we move to figure \ref{a001} (where $\alpha=0.01$, low-mixing
domain) we are able to appreciate some interplay between the two
scattering processes. In fact, in this case, while the $zz\to zz$
scattering eigenchannel allows the existence of a heavy Higgs of more
than $10$ TeV, 
%and a loose constraint applies also to the light Higgs,
the $z'z'\to z'z'$ scattering channel strongly limits the allowed mass
region, with a ``cut-off'' on the heavy Higgs mass almost insensible
to the light Higgs mass (and the value of the mixing angle, since we
are in the low-mixing domain), that is
\begin{eqnarray}\label{low-maxmh2}
{\rm{Max}}(m_{h_2})\simeq 2\sqrt{\frac{2\pi}{3}}x,
\end{eqnarray} 
which is in agreement (under different theoretical assumptions though)
with the result in \cite{Robinett:1986nw}; from a
graphical point of view, in figure \ref{a001} the (green) hollow area
represents the allowed configuration space at $x=3.5$ TeV, while at
$x=10$ TeV the allowed portion of the $m_{h_1}$-$m_{h_2}$ subspace
increases until the (green) double-lines shadowed region, finally the
constraint relaxes to the (green) single line shadowed region when $x=35$
TeV.

This interplay effect arising (somewhat unintuitively) for Higgs
low-mixing is due to the fact that the consequent decoupling between
the two Higgs states requires the light(heavy) Higgs state to
independently keep the scattering matrix elements of the
$z^{(')}z^{(')}\rightarrow z^{(')}z^{(')}$ process unitary, thus realising
two separate constraints: the first on the light (SM-like) Higgs mass 
due to the $zz\rightarrow zz$ unitarisation and the second on the
heavy ($B-L$ like) Higgs mass due to the $z'z'\rightarrow z'z'$ unitarisation.

To summarise, given a value of the singlet Higgs VEV $x$ (compatible
with experiment), the upper bound on the light Higgs boson mass varies
between the SM limit and the experimental lower limit from LEP as long
as the upper bound for the heavy
Higgs mass increases. Moreover, when $\alpha$ assumes a value included
in the high-mixing domain, the strongest bound comes from the
unitarisation of the $z$-boson scattering, whilst in the low-mixing
domain the bound on the heavy Higgs mass coming from that channel
relaxes and the unitarisation induced by the $z'$-boson scattering
becomes so important to also impose a cut-off (which depends linearly
on $x$) on the heavy Higgs mass.

This is a very important result, because it allows us to conclude
that, whichever the Higgs mixing angle, both Higgs boson masses of the
$B-L$ model are bounded from above. As examples of typical values for
the heavy Higgs mass, in table \ref{tab:mh2vsvev}, we show some upper
bounds that universally apply (i.e., no matter what the mixing angle
is) once the singlet Higgs VEV is given.

\begin{table}[!htbp]
\begin{center}
\begin{tabular}{|c|c|}
\hline
$x$ (TeV) & Max$(m_{h_2})$ (TeV) \\
\hline
$3.5$ & $\simeq 10$ \\
\hline
$7$ & $\simeq 20$ \\
\hline
$10$ & $\simeq 30$ \\
\hline
$20$ & $\simeq 60$ \\
\hline
$35$ & $\simeq 100$ \\
\hline
\end{tabular}
\end{center}
\caption{Universal upper bound on the heavy Higgs mass, $m_{h_2}$, in
  the $B-L$ model as a function of the singlet Higgs VEV, $x$.}
\label{tab:mh2vsvev}
\end{table}

Before we move on, it is also worth re-emphasising that,
if the Higgs mixing angle is such that we are in the high-mixing case,
the upper bound on the heavy Higgs boson mass coming from $z$-boson scattering
is more stringent
than the one coming from $z'$-boson scattering and it is totally
independent from the chosen singlet Higgs VEV.

Nowadays, it is important to refer in our analysis to the possibility of
a Higgs boson discovery either at Tevatron or Large Hadron Collider
(LHC). Thus, if we suppose that a light or heavy Higgs mass $m_{h_1}$
has been already measured by an experiment it is interesting to study
the $\alpha$-$m_{h_2}$ parameter space, to see whether an hitherto
unassigned Higgs state can be consistent with a minimal $B-L$
scenario.

%
%\begin{figure}[h]
%  \subfloat[]{ 
%  \label{x10_a005}
%  \includegraphics[angle=0,width=0.49\textwidth ]{Fig/mh1vsmh2_x10_a005.eps}}
%  \hspace{0.1cm}  
%\subfloat[]{ 
%  \label{x35_a035}
%  \includegraphics[angle=0,width=0.33\textwidth ]{Fig/mh1vsmh2_x35_a035.eps}}
%  \hspace{0.1cm}
%  \subfloat[]{
%  \label{x10_a021}
%  \includegraphics[angle=0,width=0.49\textwidth ]{Fig/mh1vsmh2_x10_a021.eps}}
%  \vspace*{-0.5cm}
%  \caption{Higgs bosons mass limits in the $B-L$ model coming from
%  the unitarity condition $|\textrm{Re}(a_0)|\le \frac{1}{2}$ applied
%  to the $zz \rightarrow zz$ (straight
%  line), $z'z' \rightarrow z'z'$ and $h_2h_2 \rightarrow h_2h_2$
%  (dotted lines) scattering; it has
%  been plotted for $\alpha=0.005$ (\ref{x10_a005})
%, $\alpha=0.035$ (\ref{x35_a035}) 
%  and $\alpha=0.021$ (\ref{x10_a021}) in units
%  of $\frac{\pi}{2}$, and $x=10$ TeV. The
%  shadowed region corrisponds to the limit $m_{h_2}>m_{h_1}$. The red
%  line surrounds the allowed configuration space.}
%  \label{mh1vsmh2_x10}
%\end{figure}
%
%

\begin{figure}[!ht]
  \subfloat[]{ 
  \label{x35_mh1150}
  \includegraphics[angle=0,width=0.49\textwidth ]{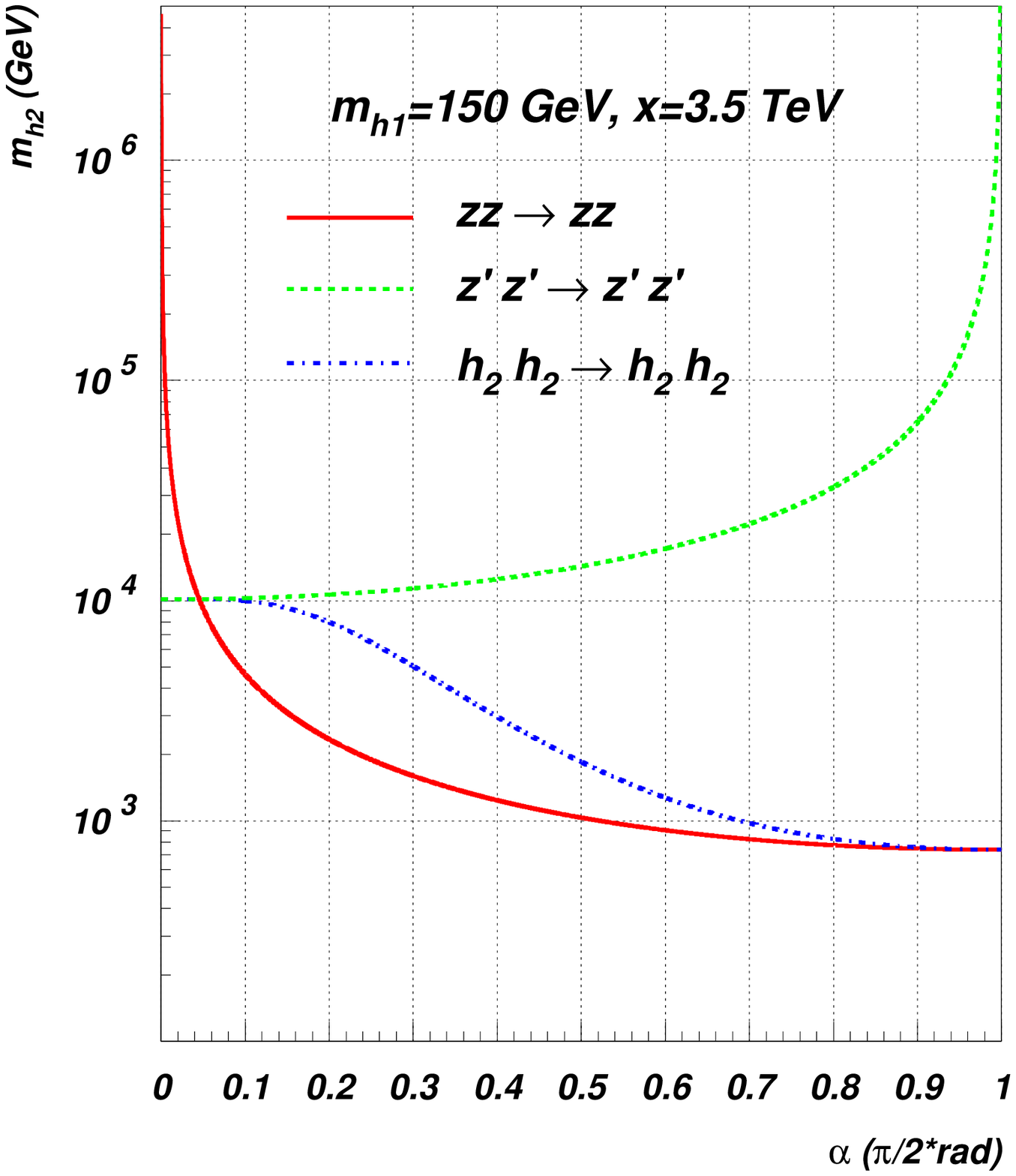}}
%  \hspace{0.1cm}  
%\subfloat[]{ 
%  \label{x35_a035}
%  \includegraphics[angle=0,width=0.33\textwidth ]{Fig/mh1vsmh2_x35_a035.eps}}
%  \hspace{0.1cm}
  \subfloat[]{
  \label{x35_mh1700}
  \includegraphics[angle=0,width=0.49\textwidth ]{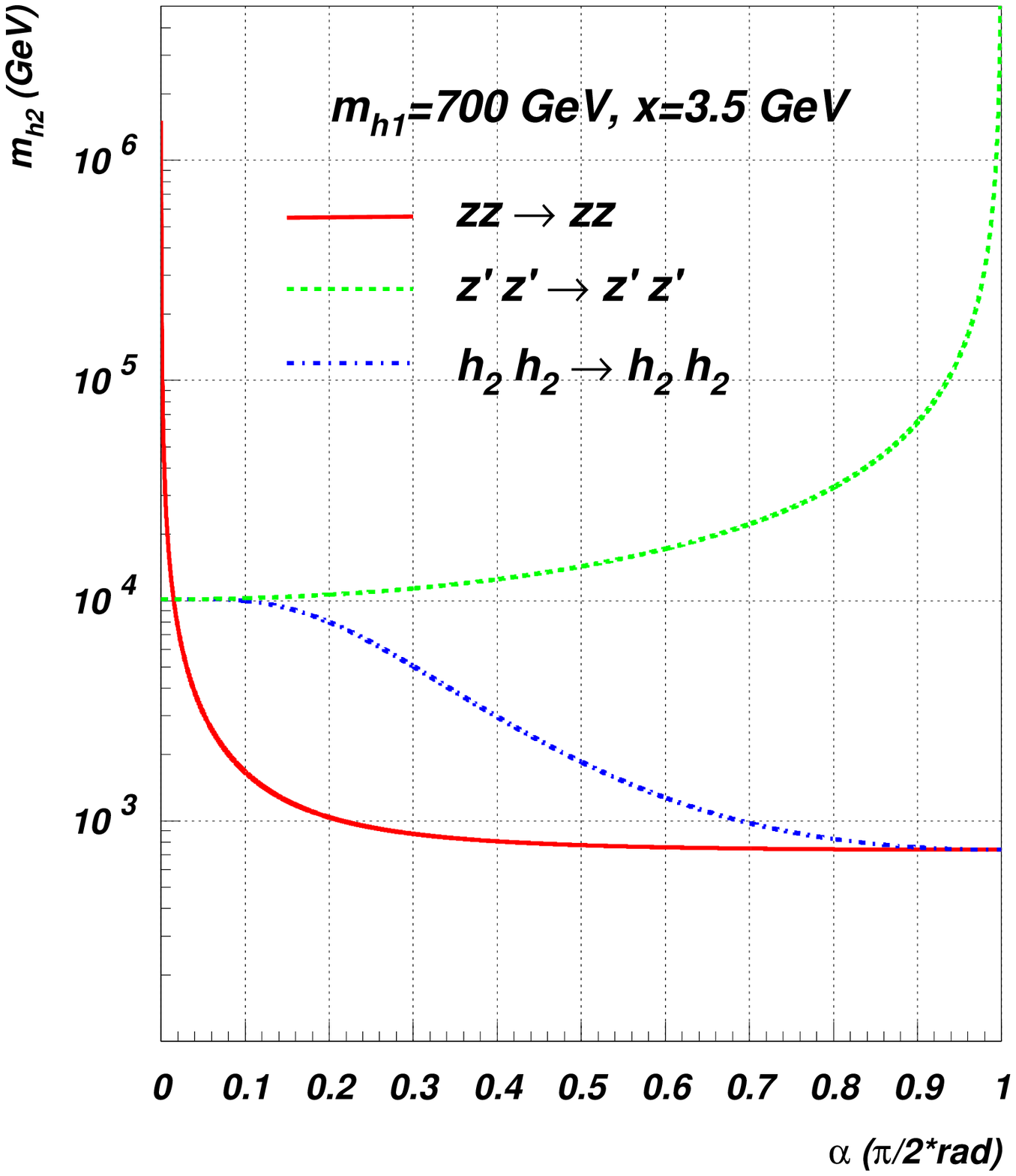}}
  \\
  \subfloat[]{ 
  \label{x350_mh1150}
  \includegraphics[angle=0,width=0.49\textwidth ]{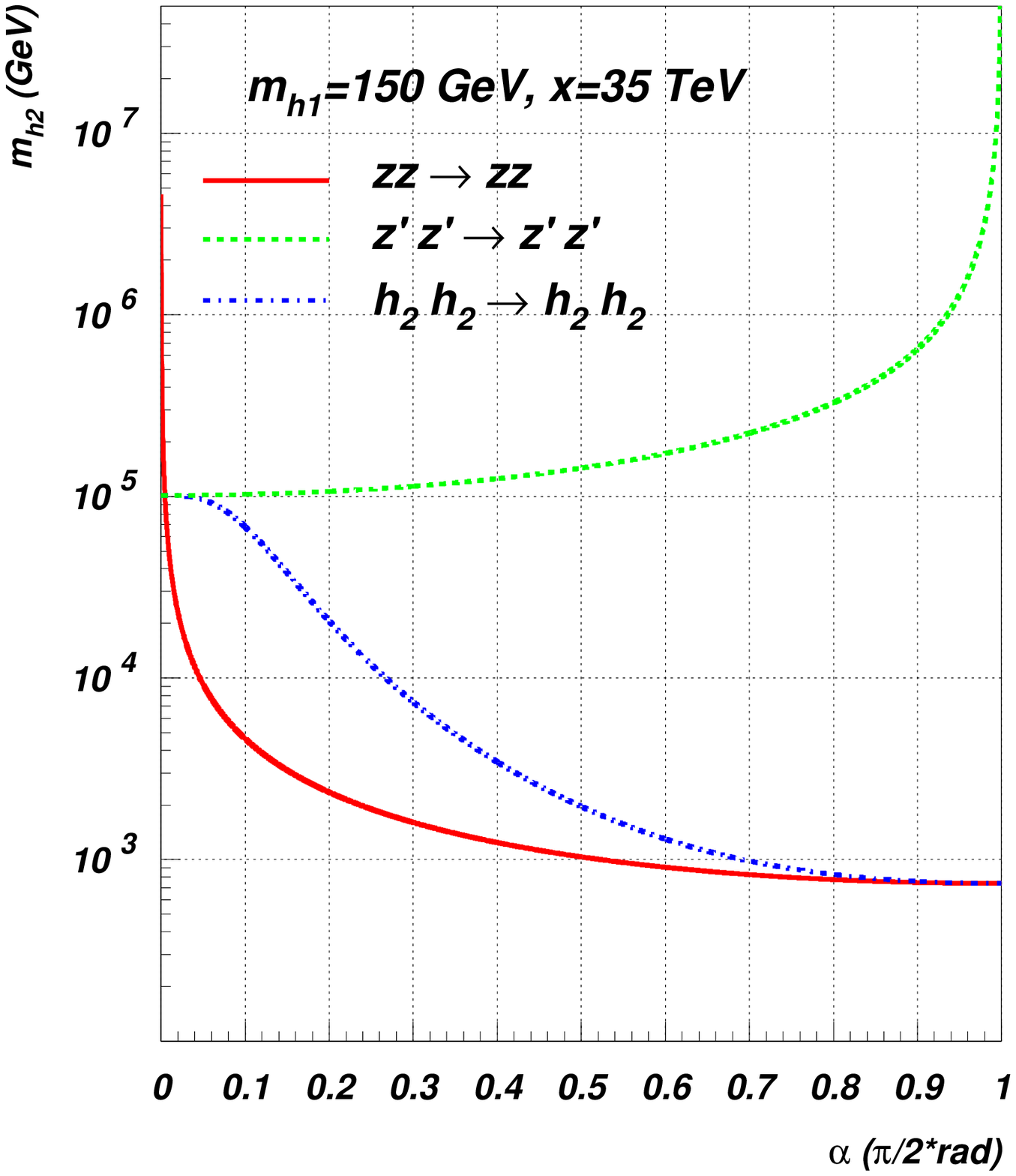}}
%  \hspace{0.1cm}  
%\subfloat[]{ 
%  \label{x35_a035}
%  \includegraphics[angle=0,width=0.33\textwidth ]{Fig/mh1vsmh2_x35_a035.eps}}
%  \hspace{0.1cm}
  \subfloat[]{
  \label{x350_mh1700}
  \includegraphics[angle=0,width=0.49\textwidth ]{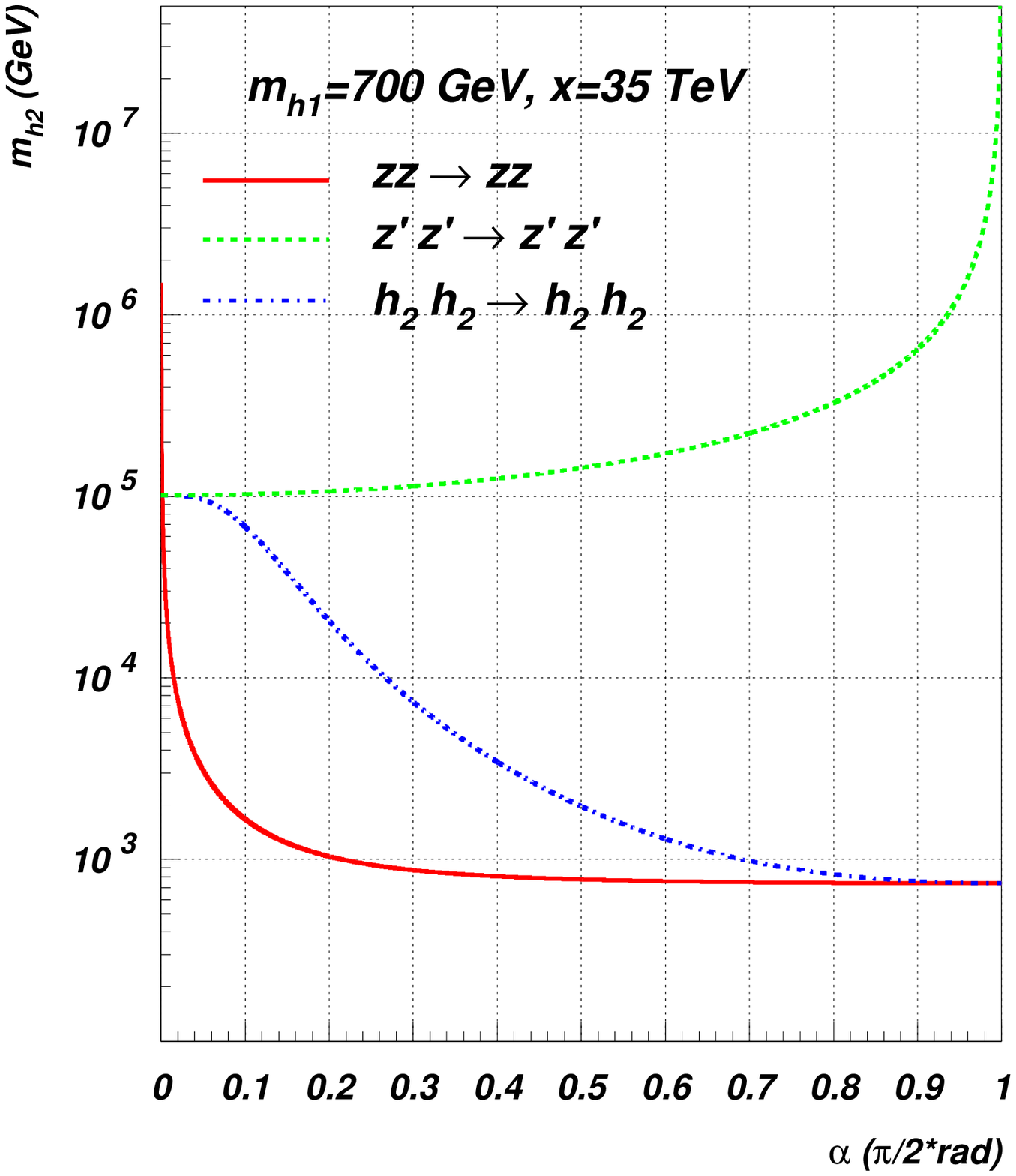}}
  \vspace*{-0.5cm}
  \caption{Heavy Higgs boson mass limits plotted against the mixing
  angle $\alpha$ in the minimal $B-L$ model. We have applied the
  unitarity condition $|\textrm{Re}(a_0)|\le \frac{1}{2}$ on $zz
  \rightarrow zz$ (red straight line), $z'z' \rightarrow z'z'$
  (green dashed line) and $h_2h_2 \rightarrow h_2h_2$ (blue
  dashed-dotted line) scatterings. This has been plotted for two fixed
  values of the light Higgs boson mass ($m_{h_1}=150$ GeV
  (\ref{x35_mh1150}, \ref{x350_mh1150}) and $m_{h_1}=700$ GeV
  (\ref{x35_mh1700}, \ref{x350_mh1700})) and of the singlet Higgs VEV
   ($x=m_{Z'}/(2g'_1)=3.5$ TeV (\ref{x35_mh1150}, \ref{x35_mh1700}) and
   $x=m_{Z'}/(2g'_1)=35$ TeV (\ref{x350_mh1150}, \ref{x350_mh1700})).}
  \label{mh2vsa}
\end{figure}

To this end, in figure \ref{mh2vsa} we fix $m_{h_1}$ and $x$ at two
extreme configurations: we take $m_{h_1}=150$ GeV as minimum value
(conservatively, taking a figure that is allowed by the experimental
lower bound established by LEP for a SM Higgs boson) and
$m_{h_1}=700$ GeV as maximum value (close to the maximum allowed by
unitarity constraints, as we saw before). Then, we take $x=3.5$ TeV as
minimum value (that is, the lower limit established by LEP data
for the existence of a $Z'$ of ${B-L}$ origin) and $m_{h_1}=35$ TeV as
maximum value (that is, one order of magnitude bigger than the smallest
VEV allowed by experiment).

Even in this case we can separate the $2$-dimensional subspace
in a
low-mixing region and a high-mixing region, as before. We can identify
the first(second) as the one in which the upper bound is established
by unitarisation through the $z'$($z$)-boson scattering. The value of
the mixing angle that separates the two regions in this case
is given by
\begin{eqnarray}
\alpha={\rm arccos}\sqrt{\frac{\left( 3 m_{h_1}^2 - 8 \pi x^2 \right)
      \alpha_W }{6 m_{h_1}^2 \alpha_W - 8 \pi x^2 \alpha_W - 8 m_{W}^2}}.
\end{eqnarray}
Once the light Higgs boson mass is fixed, we can see how the
heavy Higgs boson mass is bounded from above through the value defined 
by eq.~(\ref{low-maxmh2}) through the $z'z'\rightarrow z'z'$ channel,
and this  occurs in the low-mixing region.
In particular, we can notice how the $z'$-constraining function
reaches a plateau and overlaps with the $h_2 h_2 \rightarrow h_2h_2$
eigenchannel bound. Moreover, if we pay attention to the high-mixing
region, we see that, if $m_{h_1}$ is fixed to some low value, then
the bound on the heavy Higgs mass relaxes much more as the mixing gets
smaller and smaller with respect to the the situation in which
$m_{h_1}$ is large, where the unitarisation is shared almost
equally by $m_{h_2}$ and $m_{h_1}$.

%\newpage

%\begin{comment}
%\begin{figure}[h]
%  \subfloat[]{ 
%  \label{x350_mh1100}
%  \includegraphics[angle=0,width=0.49\textwidth ]{Fig/x35TeV_mh1100GeV.eps}}
%  \hspace{0.1cm}  
%\subfloat[]{ 
%  \label{x35_a035}
%  \includegraphics[angle=0,width=0.33\textwidth ]{Fig/mh1vsmh2_x35_a035.eps}}
%  \hspace{0.1cm}
%  \subfloat[]{
%  \label{x350_mh1700}
%  \includegraphics[angle=0,width=0.49\textwidth ]{Fig/x35TeV_mh1700GeV.eps}}
%  \vspace*{-0.5cm}
%  \caption{Heavy Higgs boson mass limits plotted against the mixing
%  angle $\alpha$ in the minimal $B-L$ model; the most stringent value
%  (red-continue line) comes from
%  the unitarity condition $|\textrm{Re}(a_0)|\le \frac{1}{2}$ applied
%  to $zz \rightarrow zz$, $z'z' \rightarrow z'z'$ and $h_2h_2
%  \rightarrow h_2h_2$ scatterings; it has
%  been plotted for two fixed values of the Light Higgs boson mass
%  ($m_{h_1}=100$ GeV (\ref{x350_mh1100}) and $m_{h_1}=700$ GeV
%  (\ref{x350_mh1700})) and $x$ ($m_{Z'}/(2g'_1)=35$ TeV).}
%  \label{mh2vsa_x350}
%\end{figure}
%\end{comment}

\subsection{Unitarity bounds at finite energy}

So far what we did was to impose the unitarity bound at the
infinite energy scale ($\sqrt s\rightarrow \infty$).
Now we can ask how this bound could change if we choose to evaluate
the same limit at some finite value of $\sqrt s$ in order to
understand if there are new configurations in the $3$-dimensional
parameter space that could invalidate our previous discussion.
For this, we study how the spherical partial wave amplitude is
modified by changing the parameter $\sqrt s$, in order to understand
if the unitarity bounds loosen somewhat.

We want to refer this kind of analysis to the $m_{h_1}$-$m_{h_2}$
subspace (at fixed $\alpha$) because it is simpler to isolate the
case in which the most stringent bound comes from the $zz\rightarrow
zz$ eigenchannel only (imposing the condition in eq.~(\ref{condition})
to the function defined by eq.~(\ref{Ta0zz})). In fact, we have already
proven that in the high energy limit, even for a small mixing angle,
say $\alpha=0.1$, we can limit our analysis to this one eigenchannel
only and plot the integrated configuration space in function of $\sqrt
s$. We will eventually demonstrate that this assumption is not
spoiled at any finite energy scale. We remind the reader that, in
order to avoid irrelevant complications, in eq.~(\ref{Ta0zz}) we
assumed that $m^2_{h_1},m^2_{h_2}\gg m^2_Z$. For this, in this
subsection we take $m^2_{h_1},m^2_{h_2}\simeq 10m^2_Z$.

\begin{figure}[!ht]
  \subfloat[]{ 
  \label{s_c_tot}
  \includegraphics[angle=0,width=0.49\textwidth ]{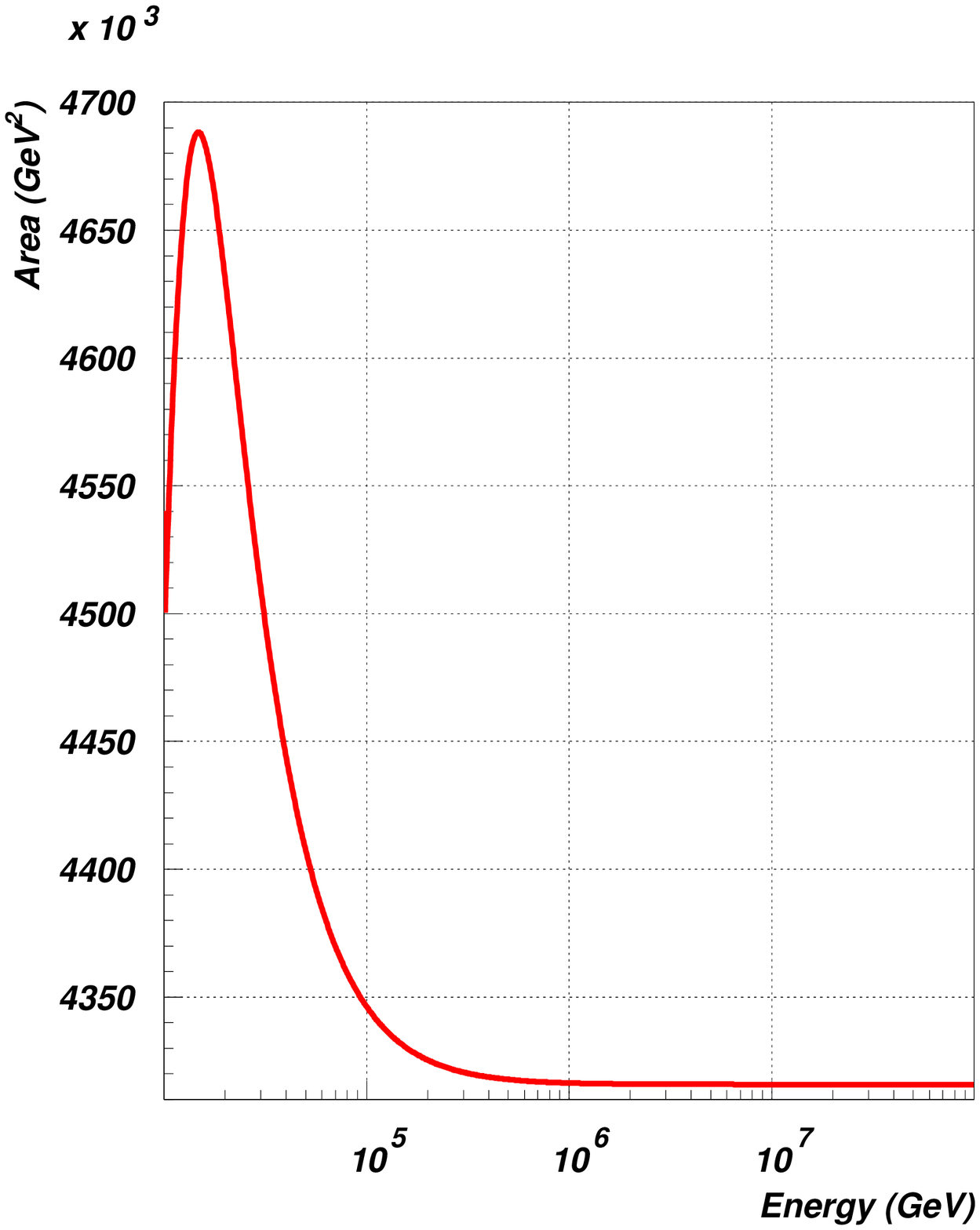}}
  \subfloat[]{
  \label{s_c_max}
  \includegraphics[angle=0,width=0.49\textwidth ]{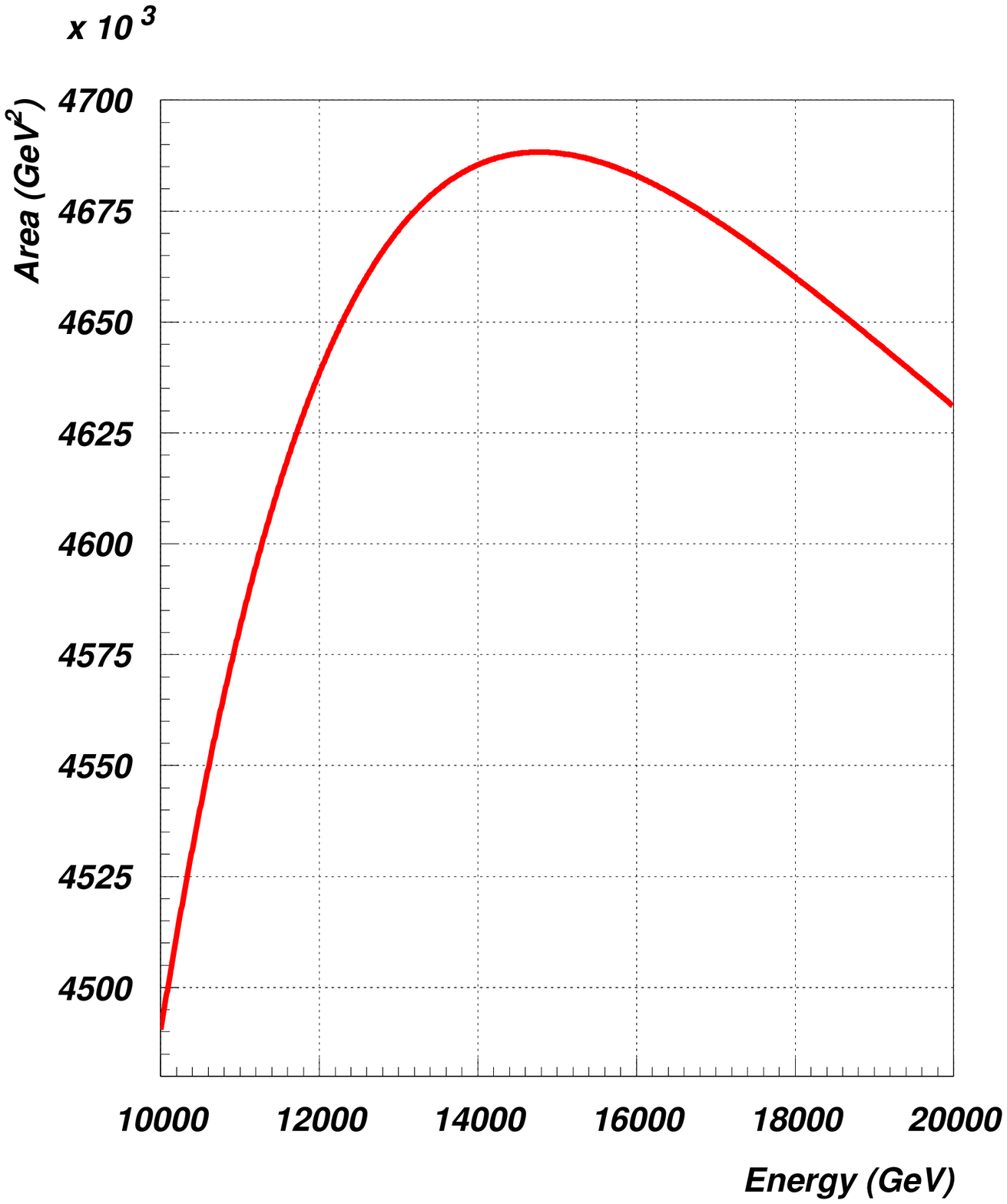}}
  \vspace*{-0.5cm}
  \caption{Integrated area of the two Higgs mass space allowed by
    the unitarity condition $|\textrm{Re}(a_0)|\le \frac{1}{2}$
    applied to the scattering channel $zz\rightarrow zz$
    plotted against the centre-of-mass energy of the process at fixed
    value of the mixing angle, $\alpha=0.1$.}
  \label{s_c}
\end{figure}

As we see from figure \ref{s_c}, the plotted function has a local
maximum,
after which it tends to the asymptotic value
established by the infinite energy limit. At the peak (corresponding
to the critical energy $\sqrt s_c \simeq 14.7$ TeV), the allowed 
configuration space at fixed angle
$\alpha=0.1$ is maximised. In short, by
comparing the allowed space in the infinite energy limit ($A_U$) and
the one in the critical configuration ($A_C$) we see that $A_C\simeq
1.09 A_U$, i.e., a mere $9\%$ difference in configuration space.

\begin{figure}[!ht]
%  \subfloat[]{ 
%  \label{UBvsTB_min}
%  \includegraphics[angle=0,width=0.49\textwidth ]{Fig/02_min_UBvsTB.eps}}
%  \subfloat[]{
\begin{center}
  \label{UBvsTB}
  \includegraphics[angle=0,width=0.74\textwidth]{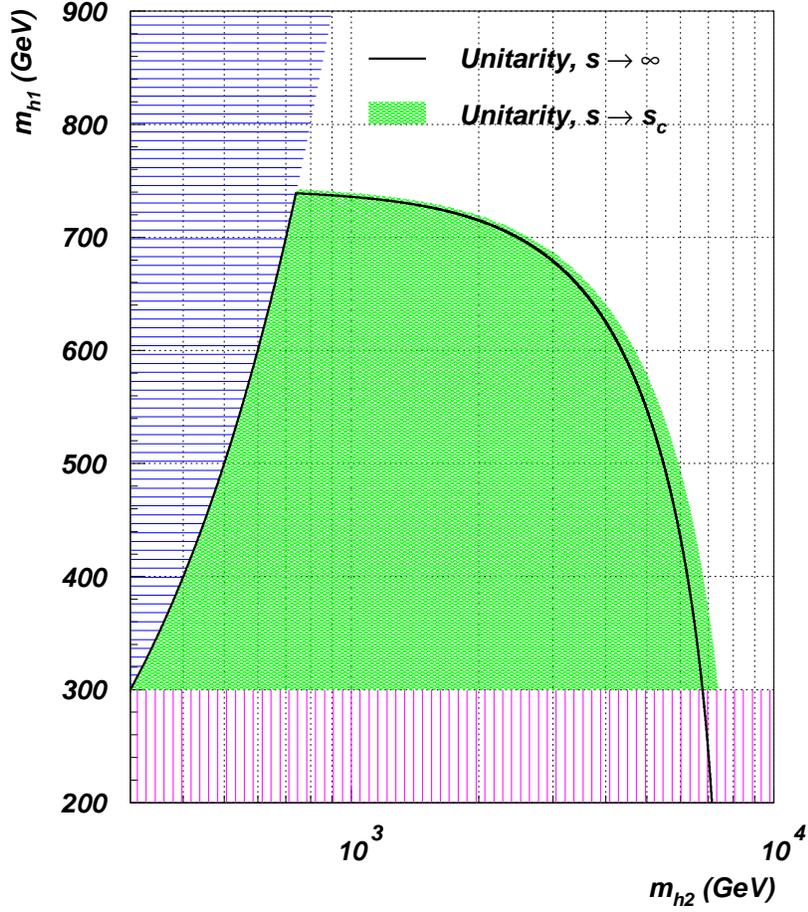}
\end{center}
%}
  \vspace*{-0.5cm}
  \caption{Higgs bosons mass limits in the $B-L$ model coming from
  the unitarity condition $|\textrm{Re}(a_0)|\le \frac{1}{2}$ applied
  to $zz \rightarrow zz$ scattering in the high energy limit
  ($\sqrt s\rightarrow \infty$, black line) and in the critical
  energy limit ($\sqrt s\rightarrow \sqrt{s_c}$, green hollow area) for a
  fixed value of the mixing angle ($\alpha=0.1$). They have been
  plotted for the local maximum of the integrated area, corresponding
  to the critical energy $\sqrt s_c \simeq 14.7$ TeV. The
  (blue) horizontal shadowed region corresponds to the unphysical
  configurations with $m_{h_1}>m_{h_2}$. The (violet) vertical
  shadowed region is excluded by the assumption
  $m^2_{h_1},m^2_{h_2}>10m^2_Z$.}
%  \label{UBvsTB}
\end{figure}

Finally, in figure 4, we want to show how
the infinite energy limit relaxes in the critical energy case.
Here, in essence, it emerges the fact that, generally, if we choose a
finite critical energy $\sqrt s_c$ instead of the infinite energy
limit and we look at
the new bounds, we cannot gain more than just a few percent
differences in the allowed mass space. In practise then, we can
conclude that, if we choose a small mixing angle, there is no
significant change in the quantitative analysis of the
$m_{h_1}$-$m_{h_2}$ space with respect to the limits obtained in the
infinite energy limit.
(Although not illustrated here, we verified that we reach the
same conclusions if we allow for large Higgs mixing instead.)

%%%%%%%%%%%%%%%%%%%%%%%%%%%%%%%%%%%%%%%%%%%%%%%%%%%%%%%%%%

\section{Conclusions}
\label{Sec:Conclusions}
In summary, we have presented a full theoretical analysis on unitarity
bounds in the Higgs sector of the minimal $B-L$ model.
The scope of this endeavour was to clarify the role of the two Higgs bosons in
the unitarisation of vector and scalar bosons scattering amplitudes,
that we know it must hold at any energy scale.

Using the equivalence theorem, we have evaluated the spherical partial
wave amplitude of all possible two-to-two scatterings in the scalar
Lagrangian at an infinite energy,
identifying the $zz \rightarrow zz$ and $z'z'\rightarrow z'z'$
processes as the 
most relevant scattering channels for this analysis ($z^{(')}$ is the
would-be Goldstone boson of the $Z^{(')}$ vector boson).

Then, we have shown that these two channels impose an upper bound on
the two Higgs masses: the light one cannot exceed the SM
bound while the limit on the heavy one is established by the
singlet Higgs VEV, whose value is presently constrained by LEP
and could shortly be extracted by experiment following a possible
discovery of a $Z'$.

We also studied how the discovery of a light Higgs boson at either
Tevatron or LHC could impact on the heavy Higgs mass bounds
in the $B-L$ model and we
discovered that the lighter the $h_1$ mass the more loose is the bound
on $m_{h_2}$, except in the low-mixing region ($\alpha \rightarrow 0$)
of the Higgs parameter space, in which the knowledge of the $x$ VEV is
again fundamental. 

Furthermore, we studied not only the infinite energy limit, but also
some lower energy critical configuration in which the Higgs mass
bounds become the most loose possible, and we discovered that in
general there are small (and not relevant) differences between the
limits obtained in the two cases, which amount to a few
percent at most.

In conclusion, in the minimal $B-L$ framework, we expect that a TeV
machine (like Tevatron, LHC or a future Linear Collider) should produce
evidence of at least a light Higgs boson. The interesting possibility
appearing in the $B-L$ model is that the companion heavy Higgs mass is
bound to be within the reach of the machine, with the actual maximum
value been dictated by the VEV $x$, i.e., the ratio between the $Z'$
mass and its coupling, both extractable by experiment.

%%%%%%%%%%%%%%%%%%%%%%%%%%%%%%%%%%%%%%%%%%%%%%%%%%%%%%%%%%

\section*{Acknowledgements} 
SM thanks Alessandro Ballestrero for helpful discussions.

%%%%%%%%%%%%%%%%%%%%%%%%%%%%%%%%%%%%%%%%%%%%%%%%%%%%%%%%%%

\appendix
%\section{Intermediate vector boson exchange in the minimal $B-L$
%  Lagrangian}
%\input{appe_b.tex}
%\label{app:b}
\section{Interaction potential of the scalar Lagrangian} 
In this appendix, we rewrite the interaction part of
eq.~(\ref{new-potential}) in terms of mass eigenstates, separating
four-point and three-point functions and classifying them by the nature
of the involved fields.

The part of the interacting potential that contains four-point
functions involving only would-be Goldstone bosons is:

\begin{eqnarray}\label{4-goldstone}
&\ &V_{4,g}=\nonumber \\
&-&\frac{\pi \alpha_W  \left(m_{h_1}^2 \cos^2{\alpha} + m_{h_2}^2 \sin^2{\alpha}
\right)}{8 m_W^2}(w^+w^-+z^2)^2
\nonumber \\
&-&\frac{(g'_1)^2 \left(m_{h_1}^2\sin^2{\alpha}
+m_{h_2}^2 \cos^2{\alpha} \right)}{2 m_{Z'}^2}(z')^4 \nonumber \\
&-&\frac{\sqrt{\pi \alpha_W}  g'_1  \left( m_{h_2}^2 - m_{h_1}^2 \right)
\sin{(2\alpha)}}{4 m_W m_{Z'}}(w^+w^-+z^2)(z')^2.
\end{eqnarray}

The part of the interacting potential that contains four-point
functions involving both would-be Goldstone and Higgs bosons is:

\begin{eqnarray}\label{4-mixed}
&\ &V_{4,hg}= \nonumber \\
&-&\frac{\sqrt{\pi \alpha _W}\cos{\alpha}}{4 m_W^2 m_{Z'}}\big[ 2 
g'_1 \left( m_{h_2}^2 - m_{h_1}^2 \right) m_W \sin^3{\alpha}
 \nonumber \\
&+& \sqrt{\pi \alpha _W}  \left( 
m_{h_1}^2 \cos^2{\alpha} + m_{h_2}^2 \sin^2{\alpha} \right) m_{Z'} \cos{\alpha}
 \big]h^2_1(w^+w^-+z^2) \nonumber \\
&-&\frac{\sqrt{\pi \alpha _W}\sin{\alpha}}{4 m_W^2 m_{Z'}}\big[ 2 
g'_1  \left( m_{h_2}^2 - m_{h_1}^2 \right)m_W \cos^3{\alpha}
 \nonumber \\
&+& \sqrt{\pi \alpha _W}  \left( 
m_{h_1}^2 \cos^2{\alpha} + m_{h_2}^2 \sin^2{\alpha} \right) m_{Z'} \sin{\alpha}
 \big]h^2_2(w^+w^-+z^2) \nonumber \\
&-&\frac{\sqrt{\pi \alpha_W}\sin{(2\alpha)}}{4 m_W^2 m_{Z'}}\big[
g'_1  \left( m_{h_2}^2 - m_{h_1}^2 \right)m_W \sin{(2\alpha)}
 \nonumber \\
&+& \sqrt{\pi \alpha _W}  \left( 
m_{h_1}^2 \cos^2{\alpha} + m_{h_2}^2 \sin^2{\alpha} \right) m_{Z'} 
 \big]h_1h_2(w^+w^-+z^2) \nonumber \\
&-&\frac{g'_1 \sin{\alpha}}{2 m_W
m^2_{Z'}}\big[ -\sqrt{\pi \alpha_W}  \left( m_{h_2}^2 - m_{h_1}^2 \right)
m_{Z'} \cos^3{\alpha} \nonumber \\
&+& 2 g'_1 \left( 
m_{h_1}^2 \sin^2{\alpha} + m_{h_2}^2 \cos^2{\alpha} \right) m_{W} \sin{\alpha}
 \big]h^2_1(z')^2 \nonumber \\
&-&\frac{g'_1 \cos{\alpha}}{2 m_W
m^2_{Z'}}\big[ -\sqrt{\pi \alpha_W}  \left( m_{h_2}^2 -
m_{h_1}^2 \right) m_{Z'} \sin^3{\alpha}
\nonumber \\
&+& 2 g'_1 \left( 
m_{h_1}^2 \sin^2{\alpha} + m_{h_2}^2 \cos^2{\alpha} \right) m_{W} \cos{\alpha}
 \big]h^2_2(z')^2 \nonumber \\
&-&\frac{g'_1\sin{(2\alpha)}}{4 m_W m^2_{Z'}}\big[
\sqrt{\pi \alpha_W}  \left( m_{h_2}^2 - m_{h_1}^2 \right) m_{Z'} \sin{(2\alpha)}
\nonumber \\
&-& 4 g'_1 \left( 
m_{h_1}^2 \sin^2{\alpha} + m_{h_2}^2 \cos^2{\alpha} \right) m_{W} 
 \big]h_1h_2(z')^2.
\end{eqnarray}
%%%%%%%%%%%%%%%%%%%%%%%%%%%%%%%%%%%%%%%%%%%%%%%%%%%%%%%5

The part of the interacting potential that contains four-point
functions involving only Higgs bosons is:

\begin{eqnarray}\label{4-higgs}
&\ &V_{4,h}= \nonumber \\
&-&\frac{1}{16} \Bigg[ \frac{8  (g'_1)^2 \left(
m_{h_1}^2 \sin^2{\alpha} +
m_{h_2}^2 \cos^2{\alpha} \right) \sin^4{\alpha}}{m_{Z'}^2} \nonumber \\
&+&\frac{\sqrt{\pi \alpha_W}  g'_1  \left( m_{h_2}^2 - m_{h_1}^2 \right)
\sin^3{(2\alpha)}}{m_W m_{Z'}} \nonumber \\
&+& \frac{2 \pi \alpha_W  \left(
m_{h_1}^2\cos^2{\alpha} + m_{h_2}^2 \sin^2{\alpha}\right) \cos^4{\alpha}}{m_W^2}\Bigg]
h^4_1 \nonumber \\
&-&\frac{\sin{(2\alpha)}}{4 m_W^2 m_{Z'}^2} \left( 2 g'_1 m_W \sin{\alpha}
+\sqrt{\pi \alpha_W} 
m_{Z'} \cos{\alpha} \right) \times \nonumber \\
&\times &\Big[ -2 
g'_1 \left(m_{h_1}^2\sin^2{\alpha} + m_{h_2}^2 \cos^2{\alpha} \right)
m_W \sin{\alpha}
\nonumber \\
&+& \sqrt{\pi \alpha_W} \left(m_{h_1}^2\cos^2{\alpha} + m_{h_2}^2 \sin^2{\alpha} \right)
m_{Z'} \cos{\alpha} \Big]h_1^3h_2
\nonumber \\
&-&\frac{\sin{(2\alpha)}}{16 m_W^2
m_{Z'}^2} \Big[ 12  (g'_1)^2 \left( 
m_{h_1}^2 \sin^2{\alpha} + m_{h_2}^2 \cos^2{\alpha} \right)
m_W^2 \sin{(2\alpha)} \nonumber \\
&+&\sqrt{\pi \alpha_W} g'_1  \left( m_{h_2}^2 - m_{h_1}^2 \right) m_W
m_{Z'} (1 + 3 \cos{(4\alpha)}) \nonumber \\
&+& 3 \pi \alpha_W  \left(
m_{h_1}^2 \cos^2{\alpha} + m_{h_2}^2 \sin^2{\alpha} \right)
m_{Z'}^2 \sin{(2\alpha)} \Big] h_1^2h_2^2 \nonumber \\
&-&\frac{\sin{(2\alpha)}}{4 m_W^2 m_{Z'}^2}
\left( 2 g'_1 m_W \cos{\alpha}
+\sqrt{\pi \alpha_W} 
m_{Z'} \sin{\alpha} \right) \times \nonumber \\
&\times &\Big[-2 
g'_1 \left(m_{h_1}^2\sin^2{\alpha} + m_{h_2}^2 \cos^2{\alpha} \right)
m_W \cos{\alpha}
\nonumber \\
&+& \sqrt{\pi \alpha_W} \left(m_{h_1}^2\cos^2{\alpha} + m_{h_2}^2 \sin^2{\alpha} \right)
m_{Z'} \sin{\alpha} \Big]h_1h_2^3
\nonumber \\
&-&\frac{1}{16} \Bigg[ \frac{8  (g'_1)^2 \left(
m_{h_1}^2 \sin^2{\alpha} +
m_{h_2}^2 \cos^2{\alpha} \right)\cos^4{\alpha}}{m_{Z'}^2} \nonumber \\
&+&\frac{\sqrt{\pi \alpha_W}  g'_1  \left( m_{h_2}^2 - m_{h_1}^2 \right)
\sin^3{(2\alpha)}}{m_W m_{Z'}} \nonumber \\
&+& \frac{2 \pi \alpha_W  \left(
m_{h_1}^2\cos^2{\alpha} + m_{h_2}^2 \sin^2{\alpha}\right) \sin^4{\alpha}}{m_W^2}\Bigg]
h^4_2.
\end{eqnarray}
%%%%%%%%%%%%%%%%%%%%%%%%%%%%%%%%%%%%%%%%%%%%%%%%%%%%%%%%%%%%%%%

The part of the interacting potential that contains three-point
functions involving both would-be Goldstone and Higgs bosons is:

\begin{eqnarray}\label{3-mixed}
&\ &V_{3,hg}= \nonumber \\
&-&\frac{\sqrt{\pi \alpha_W} m_{h_1}^2 \cos{\alpha}}{2
m_W}h_1(w^+w^-+z^2) \nonumber \\
&-&\frac{\sqrt{\pi \alpha_W} m_{h_2}^2 \sin{\alpha}}{2
m_W}h_2(w^+w^-+z^2) \nonumber \\
&+&\frac{ g'_1 m_{h_1}^2 \sin{\alpha}}{m_{Z'}} h_1(z')^2 -\frac{ g'_1
m_{h_2}^2 \cos{\alpha}}{m_{Z'}} h_2(z')^2.
\end{eqnarray}
%%%%%%%%%%%%%%%%%%%%%%%%%%%%%%%%%%%%%%%%%%%%%%%%%%%%%%%%%%%%%%%5

The part of the interacting potential that contains three-point
functions involving only Higgs bosons is:
\begin{eqnarray}\label{3-higgs}
&\ & V_{3,h}= \nonumber \\
&-& \frac{m_{h_1}^2}{2}\Bigg( 
-\frac{2 g'_1 \sin^3{\alpha}}{m_{Z'}}
+\frac{\sqrt{\pi \alpha_W}\cos^3{\alpha}}{m_W}
\Bigg) h_1^3 \nonumber \\
&-&\frac{\sin{(2\alpha)}}{4 m_W m_{Z'}}
( 2 m_{h_1}^2 + m_{h_2}^2 ) \Big( 2 g'_1 m_W \sin{\alpha}
+ \sqrt{\pi \alpha_W} m_{Z'} \cos{\alpha} \Big) h_1^2h_2 \nonumber \\
&-& \frac{\sin{(2\alpha)}}{4 m_W m_{Z'}}
( m_{h_1}^2 + 2 m_{h_2}^2 ) \Big( -2 g'_1 m_W \cos{\alpha}
+ \sqrt{\pi \alpha_W} m_{Z'} \sin{\alpha}
  \Big) h_1h_2^2 \nonumber \\
&-& \frac{m_{h_2}^2}{2} \Bigg( 
\frac{2 g'_1 \cos^3{\alpha}}{m_{Z'}}
+\frac{\sqrt{\pi \alpha_W}\sin^3{\alpha}}{m_W}
\Bigg) h_2^3.
\end{eqnarray}

\label{app:a}

%%%%%%%%%%%%%%%%%%%%%%%%%%%%%%%%%%%%%%%%%%%%%%%%%%%%%%%%%%

\end{document}